# Device-to-Device Communication in Cellular Networks: A Survey

Pimmy Gandotra, *Student Member, IEEE*, Rakesh Kumar Jha, *Member, IEEE*

*Abstract*— A constant need to increase the network capacity for meeting the growing demands of the subscribers has led to the evolution of cellular communication networks from the first generation (1G) to the fifth generation (5G). There will be billions of connected devices in the near future. Such a large number of connections are expected to be heterogeneous in nature, demanding higher data rates, lesser delays, enhanced system capacity and superior throughput. The available spectrum resources are limited and need to be flexibly used by the mobile network operators (MNOs) to cope with the rising demands. An emerging facilitator of the upcoming high data rate demanding next generation networks (NGNs) is device-to-device (D2D) communication. An extensive survey on device-to-device (D2D) communication has been presented in this paper, including the plus points it offers; the key open issues associated with it like peer discovery, resource allocation etc, demanding special attention of the research community; some of its integrant technologies like millimeter wave D2D (mmWave), ultra dense networks (UDNs), cognitive D2D, handover procedure in D2D and its numerous use cases. Architecture is suggested aiming to fulfill all the subscriber demands in an optimal manner. The Appendix mentions some ongoing standardization activities and research projects of D2D communication.

## I. INTRODUCTION

Today the number of hand-held devices is drastically increasing, with a rising demand for higher data rate applications. In order to meet the needs of the next generation applications, the present data rates need a refinement. The fifth generation (5G) networks are expected and will have to fulfill these rising demands. A competent technology of the next generation networks (NGNs) is Device-to-Device (D2D) Communication, which is expected to play an indispensable role in the approaching era of wireless communication. The use of D2D communication did not gain much importance in the previous generations of wireless communication, but in 5G networks, it is expected to be a vital part. The rising trends [1] pave way for this emerging technology. With the introduction of device-to-device (D2D) communication, direct transmission between devices is possible. This is expected to improve the reliability of the link between the devices, enhance spectral efficiency and system capacity [2], with reduced latency within the networks. Such a technique is essential for fulfilling the chief goals of the mobile network operators (MNOs).

D2D communication allows communication between two devices, without the participation of the Base Station (BS), or the evolved NodeB (eNB). Proximate devices can directly communicate with each other by establishing direct links. Due to the small distance between the D2D users, it supports power saving within the network, which is not possible in case of conventional cellular communication. It promises improvement in energy efficiency, throughput and delay. It has the potential to effectively offload traffic from the network core. Hence, it is a very flexible technique of communication, within the cellular networks.

Qualcomm's FlashLinQ [3] was the first endeavor towards the implementation of device to device (D2D) communication in cellular networks. It takes advantage of orthogonal frequency division multiple access (OFDMA) in conjunction with distributed scheduling for peer discovery, link management and synchronization of timings. Another organization involved in examining D2D communication in cellular networks is 3GPP (Third Generation Partnership Project) [4], [5], [6]. D2D communication is under investigation by the 3GPP as Proximity Services (ProSe). It is expected to function as a public safety network feature in Release 12 of 3GPP. The task of standardization of device-to-device communication and the ongoing projects are briefly discussed in APPENDIX A and B. A next generation network scenario, supporting device-to-device (D2D) communication along with some general use cases is depicted in Fig.1. The most popular use cases of D2D include public safety services, cellular offloading, vehicle-to-vehicle (V2V) communication, content distribution.

In spite of the numerous benefits offered by device-to-device (D2D) communication, a number of concerns are involved with its implementation. When sharing the same resources, interference between the cellular users and D2D users needs to be controlled. For this, numerous interference management algorithms have been proposed in literature. Other concerns include peer discovery and mode selection, power control for the devices, radio resource allocation and security of the communication.

### A. Contributions:

Existing surveys [44], [47] on device-to-device (D2D) communication provide an extensive literature on the various issues in D2D communication. The authors in [44] comprehensively describe the state-of-the-art research work

Pimmy Gandotra is a M.Tech Research Scholar in School of Electronics and Communication Engineering, Shri Mata Vaishno Devi University, J&K, India. (E-mail: 14mmc015@smvdu.ac.in).
Rakesh Kumar Jha, is Assistant Professor in School of Electronics and Communication Engineering, Shri Mata Vaishno Devi University, J&K, India. (E-mail: jharakesh.45@gmail.com)

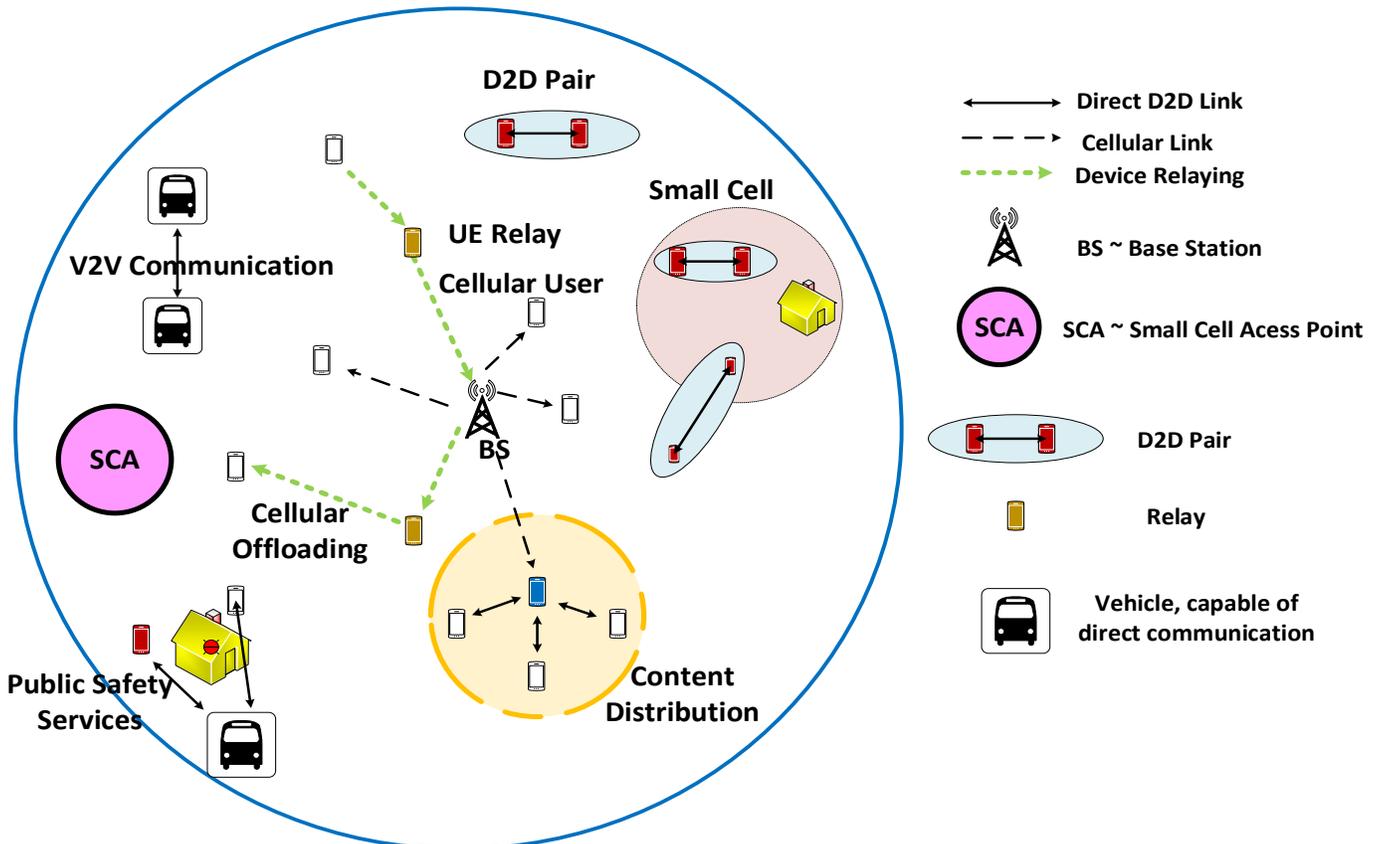

Fig.1. A General Scenario supporting device-to-device (D2D) communication

on D2D communication in LTE-Advanced networks. In [47], the literature available on D2D communication is presented as Inband D2D and Outband D2D. This survey, on the other hand, draws upon the growing need for switching towards the device-to-device (D2D) technology. Architecture for device-to-device (D2D) communication has been proposed, which clearly depicts the scenario of the next generation networks (NGNs) and is the prime focus of this survey. It aims to aid the cellular networks in near future by allocating resources optimally to the D2D users in the network and the cellular users as well, with the use of sectored antennas at the base station (BS). Such architecture has the potential to efficiently serve the rising demands of the subscribers and meet the requirements of the network operators. Additionally, a mathematical analysis has been discussed, which is the basis of any resource allocation technique, for analyzing network throughput. Number of features can be integrated with D2D communication, to enhance its utility in existing cellular systems. These have been discussed in this survey. A number of challenges exist, pertaining to the implementation of device-to-device (D2D) communication. Few important algorithms in relation to these issues have been discussed. Thus, focus of this survey is to brief about different aspects of D2D communication.

The organization of the survey is as follows: Following the introduction, a roadmap to D2D communication has been presented in Section II. An overview of device-to-device (D2D) communication has been presented in Section III. The various features which can be integrated with D2D communication to further enhance their utility and performance in cellular networks are discussed in Section IV. Incorporating D2D communication in existing cellular networks engenders a number of challenges, which have been discussed in Section V. Architecture has been proposed in this section, to overcome the issue of radio resource management. Since the architecture uses a sectored antenna at the base station, interference between D2D users and cellular users in the networks is overcome to a large extent. In the next generation networks, a number of applications are expected to be supported by D2D communication, and are discussed in Section VI. Lastly, the paper concludes in Section VII.

## II. THE ROADMAP TO DEVICE-TO-DEVICE (D2D) COMMUNICATION

Telegraphy was demonstrated by Joseph Henry and Samuel F.B. Morse, in 1832. In 1864, James Clerk Maxwell postulated wireless propagation, which was verified and demonstrated by Heinrich Hertz in 1880 and 1887, respectively. Marconi and Popov started experiments with the radio-telegraph shortly thereafter, and Marconi patented a complete wireless system in 1897. These marked the advent of wireless communication.

The very first wireless networks were discovered during the pre-industrial age. These primarily were based on Line of Sight (LOS) transmissions. These networks were then replaced by the telegraph, and later by the telephone. After the invention of the telephone, Marconi demonstrated the first radio transmission. Thereafter, radio technology rapidly gained importance as transmissions over very large distances was possible with better quality, low cost and less power.

TABLE I
COMPARISON OF D2D WITH WLAN AND BLUETOOTH

| Feature Considered | BLUETOOTH | WLAN | D2D COMMUNICATION |
|---|---|---|---|
| Pairing | Require manual pairing | Require user defined settings for access points | Base station assisted or device assisted |
| Quality of Service (QoS) | No hard QoS guarantee | No hard QoS guarantee | Provides hard QoS guarantees |
| Spectrum | Unlicensed | Unlicensed | Licensed, Unlicensed |
| Standardization | Bluetooth SIG | IEEE 802.11 | 3GPP Release 12 |
| Maximum Data rate | 25Mb/s | 54Mb/s | 5-10 Gb/s |
| Modulation Technique | GFSK | DSSS | SC-FDMA (Downlink), OFDMA (Uplink) |
| Max.Transmission Distance | 10-100 m | 32 m | Up to 500 m |
| Forward Error Correction | ARQ, FEC (MAC) | ARQ, FEC (PHY) | Low Density Parity Check codes (LDPC) |
| Max Transmit Power | 4dBm | 15dBm | 24dBm |
| Pricing | Free of cost | Free of cost | Operator decides the cost |

Initially, only analog data transmission took place. Then there was a shift towards digital data transmission. The generations of wireless networks have evolved from first generation (1G) to fifth generation (5G). A brief overview of the generations in connection to D2D communication has been given in this section.

*A. First Generation (1G)*

This generation of wireless communication came into existence in the early 1980s and supported data rates up to 2.8Kbps. These networks were circuit switched. The analog cellular technology was referred to as Analog Mobile Phone Service (AMPS) and it used Frequency Division Multiplexing (FDM). These were completely insecure networks and required large power consumption. Also, quality of calls was very low. Due to less number of subscribers during this era, the need for direct transmission was never felt.

*B. Second Generation (2G)*

It is digital cellular. It came into existence in the late 1990s. The first second generation (2G) system was Global System for Mobile (GSM). It supported a maximum data rate of up to 64kbps. Other technologies included in it are Code Division Multiple Access (CDMA) and IS-95. It provided the services like email and short message service (SMS) [7], [8]. These networks are more secure against eavesdropping, as compared to the 1G network. This generation could not handle complex data like videos.

Between 2G and 3G came another generation, 2.5G. Data rates of up to 200kbps were supported in 2.5G. Technologies included General Packet Radio Service (GPRS) and Enhanced Data Rate for GSM Evolution (EDGE). No direct communication was used in wireless communication till this period.

*C. Third Generation (3G)*

Data rates supported by the third generation networks are up to 2Mbps. These came in late 2000 and support services with improved voice quality and help maintain better Quality of Service (QoS). The technologies supported by 3G include Wideband Code Division Multiple Access (WCDMA), Universal Mobile Telecommunication System (UMTS), and Code Division Multiple Access (CDMA) 2000. Technologies like Evolution-Data Optimized (EVDO), High Speed Uplink/Downlink Packet Access (HSUPA/HSDPA) form a part of 3.5G and provide improved data rates in comparison to 3G. Though 3G is more advantageous than 2G, it requires more power than 2G networks and is costlier than 2G in terms of the plans it offers. In this generation, WLAN and Bluetooth gained popularity and allowed direct communication between devices. These techniques function in the unlicensed band, like in the industrial, scientific and medical (ISM) band, not meeting the Quality of Service (QoS) requirements of the network efficiently. Licensed band is more capable of handling the interference issue, thereby meeting the QoS needs of the cellular networks. As a result, interference management is possible with the help of a central controlling entity in the network (the base station), with D2D communication underlaying cellular communication. thus, D2D communication in cellular networks was introduced in the next generations. A comparison of Bluetooth and WLAN technologies with D2D communication has been shown in Table I.

*D. Fourth Generation (4G)*

Further enhancement in data rates are provided by 4G networks. These provide a system completely based on internet protocol (IP). Applications supported by 4G networks include Multimedia Messaging Service (MMS), Digital Video Broadcasting (DVB), HDTV, video chatting etc. Technologies include Long Term Evolution Advanced (LTE-A) and Mobile Worldwide Interoperability for Microwave Access (WiMAX). 4G networks are referred to as *MAGIC*: **M**obile multimedia, **A**nytime anywhere, **G**lobal Mobility Support, **I**ntegrated wireless solution and **C**ustomized Personal Service. Long Term Evolution- Advanced (LTE-A) introduced device-to-device (D2D) communication in cellular networks.

*E. Fifth Generation (5G)*

The fifth generation (5G) of wireless communication is the next generation networks. 4G systems will soon be replaced by 5G in order to fulfill the increasing demands of the subscribers for higher data rates and support numerous applications. It includes various enhanced technologies like Beam Division Multiple Access (BDMA) and Non- and quasi-orthogonal or Filter Bank multi carrier (FBMC) multiple access. 5G is the result of an aggregation of numerous technologies like, mmWave communication, Massive MIMO, Cognitive Radio Networks (CRNs), Visible light communication (VLC). The first four generations were completely dependent upon the base station (BS), thus called network centric. But 5G is heading towards device-centric approach, i.e. network setup and managed by the devices themselves. Device-to-Device (D2D) Communication is being considered as an essential component of the 5G networks. It is expected to result in an enhanced system capacity, increased spectral efficiency, better throughput and reduced latency. An overview of the eras of wireless communication and the services supported by them is depicted in Fig. 2. A detailed overview of the evolution of generations of wireless communication has been given in [9].

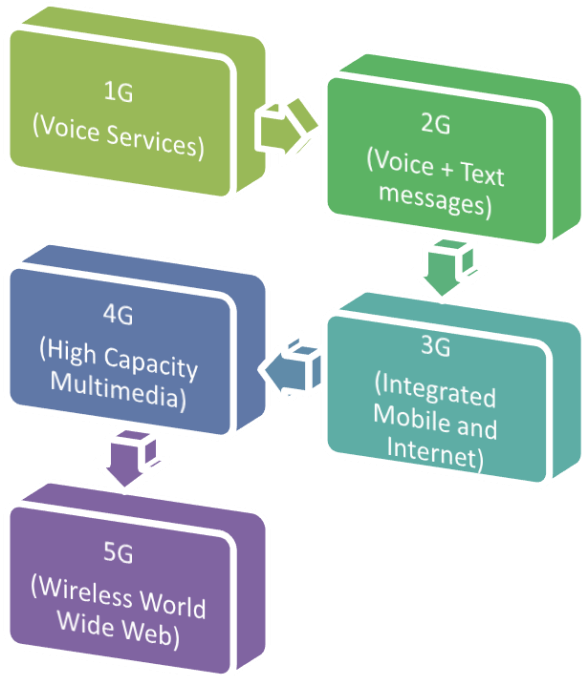

Fig.2. Generations of wireless communication

There has been a drastic growth in traffic over the years, and will continue in the years to come, as depicted in Fig. 3 [10]. This results in overloading at the base station (BS). Due to this mounting load on the base station (BS), there is an increase in the demand for power. To overcome this need for high power, some amount of traffic needs to be offloaded from the base station and here D2D communication plays a crucial role. Since D2D communication allows devices to communicate with each other without traversing the base station, load on the base station is highly reduced.

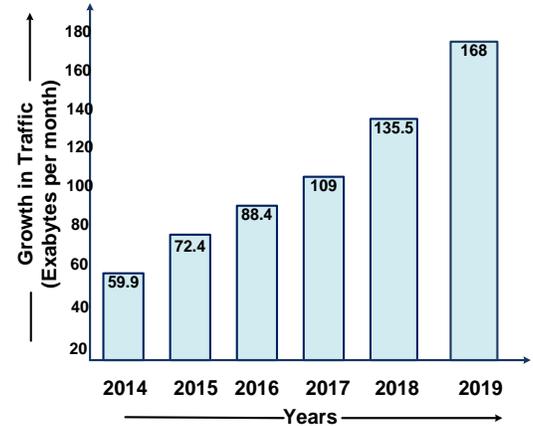

Fig.3. Rising Trends in D2D [10]

### III. OUTLINE OF DEVICE-TO-DEVICE (D2D) COMMUNICAION

Taking into consideration the architecture perspective, D2D communication networks appear to be similar to Mobile Ad-hoc Networks (MANETs) and Cognitive Radio Network (CRN). MANETs is a collection of mobile nodes which form a temporary network without the assistance of any centralized administrator. These are generally multi hop networks. A number of challenges are faced by MANETs which prevent them from providing the required Quality of Service (QoS) guarantees. These challenges include unreliable wireless channel, contention of the wireless channel, and lack of a centralized control. The nodes in MANETs suffer from severe resource constraints. In case of cognitive radio networks (CRNs), spectrum sensing is a big challenge [11]. The CR physical layer aspects also have to be addressed, in order to exploit its utility completely. Other issues in CRNs include white space detection, collision avoidance, and synchronization.

In comparison to MANETs and CRNs, D2D networks can be either base station (BS) controlled or device controlled. When base station controlled, issues in MANETs and CRN can be overcome by D2D communication. The users in cognitive radio networks (CRNs) [12] are identified as primary or secondary, which is possible in D2D communication also. But, cognitive sensing and autonomous functioning of CRNs is not supported by D2D communication. The challenges of QoS provisioning existing in MANETs are overcome by D2D communication. A brief comparison of D2D communication and MANETs has been depicted in Table II. Due to the benefits offered, device-to-device (D2D) communication is being looked upon as an effective technique to meet the rising user demands. It supports development of new applications and data offloading which is a significant contribution of this technique.

TABLE II
MANETS V/S D2D COMMUNIACATION

| *MANETs* | *D2D Communication* |
| --- | --- |
| Multi-hop Networks | One-hop networks |
| No QoS guarantee | QoS guarantee |
| No improvement in Spectral efficiency supported | Improvement in spectral efficiency supported |
| Less security guarantee | Better security guarantee |
| No centralized control | Centrally control by the base station (either fully or partially) |
| Manual connectivity | Seamless association, subject to fulfillment of distance constraint |
| No handover phenomenon | Handover phenomenon is possible |
| Poor Resource utilization; power constraint based resource utilization | Efficient resource utilization |

The fifth generation (5G) cellular networks, with Device-to-Device (D2D) Communication enabled within is considered as two-tier networks. The two tiers in these networks are referred to as the macro cell tier and the device tier. Conventional cellular communication is supported by the macro cell tier, while D2D communication is supported by the device tier. These cellular networks thus are similar to the existing networks. The difference lies in the fact that faithful services can be achieved by the devices at the cell edges and those in the congested areas within the cell. As devices in the device tier allow direct D2D communication, the base station may have a partial control or a full control over the communication between the devices. Thus, device to device (D2D) communication in the device tier is categorized into four different types [13]:

*(1) Device relaying with controlled link establishment from the operator*

Devices at the cell edges or in poor coverage areas are capable of communicating with the base station (BS) by relaying information through other devices. All tasks of establishing the communication between the devices are handled by the base station (BS). The battery life of the devices is enhanced this way. The architecture is as shown in Fig.4.

*(2) Direct communication between devices with controlled link establishment by the operator*

Two devices communicate directly with each other, with control links provided by the base station. Though direct, the communication is entirely managed by the base station. Since in (1) and (2), a central controlling entity, i.e. the base station (BS) is present, interference management is possible. The architecture is as shown in Fig.5.

*(3) Device relaying with controlled link establishment from the device*

Two devices communicate via relays, within the cellular networks. Resource allocation, setting up of call, interference management, all is managed by the devices themselves, in a distributive fashion. Control of the base station is missing. The architecture is as shown in Fig.6.

*(4) Direct communication between devices (Direct D2D) with controlled link establishment by the device*

Devices communicate directly, without aid from the base station (BS). Call setup and management are handled by the devices themselves, as in (3). The architecture is as shown in Fig.7.

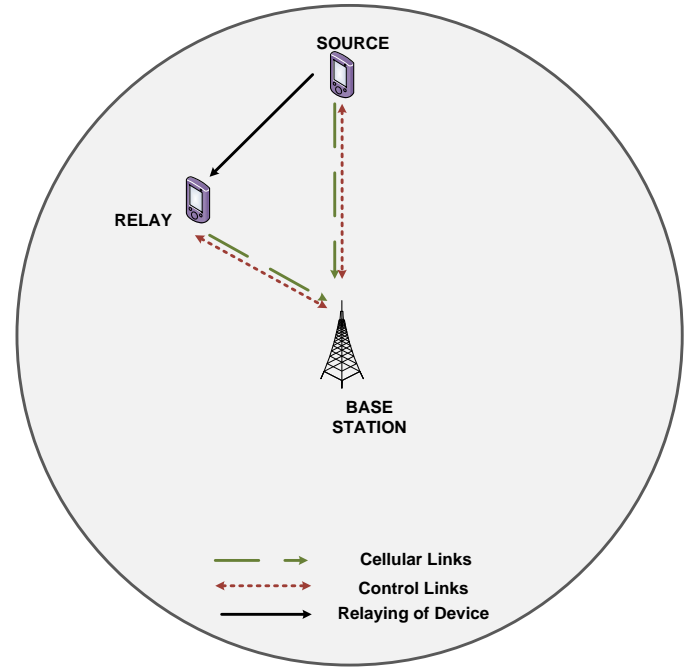

Fig.4. Relaying Devices with controlled link establishment from the operator

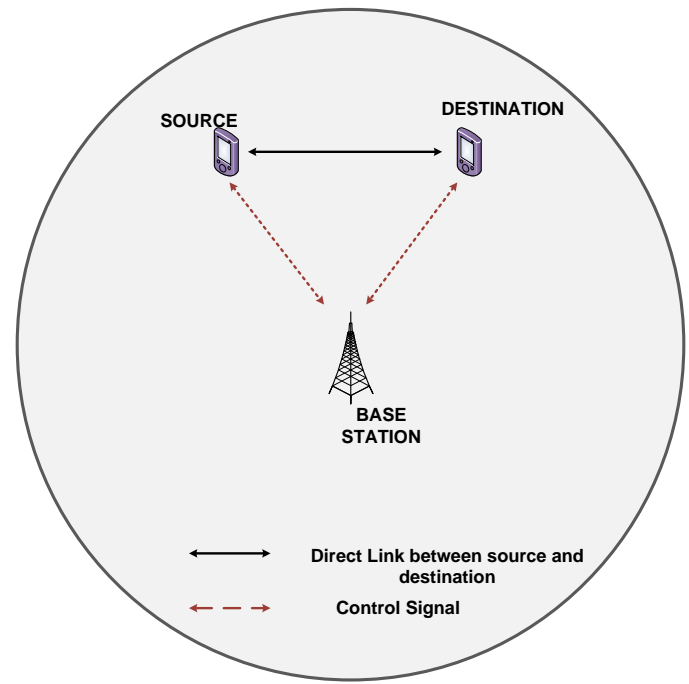

Fig.5. Direct communication between devices with operator controlled link establishment

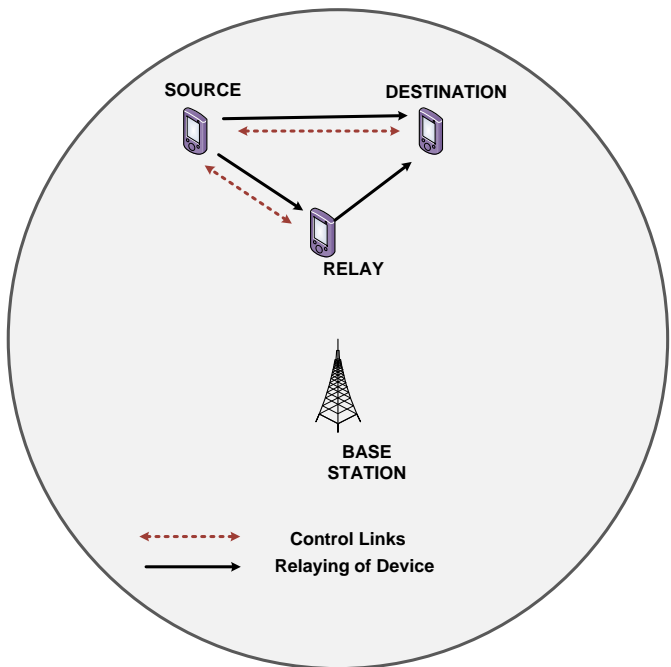

Fig.6. Relaying device with device controlled link establishment

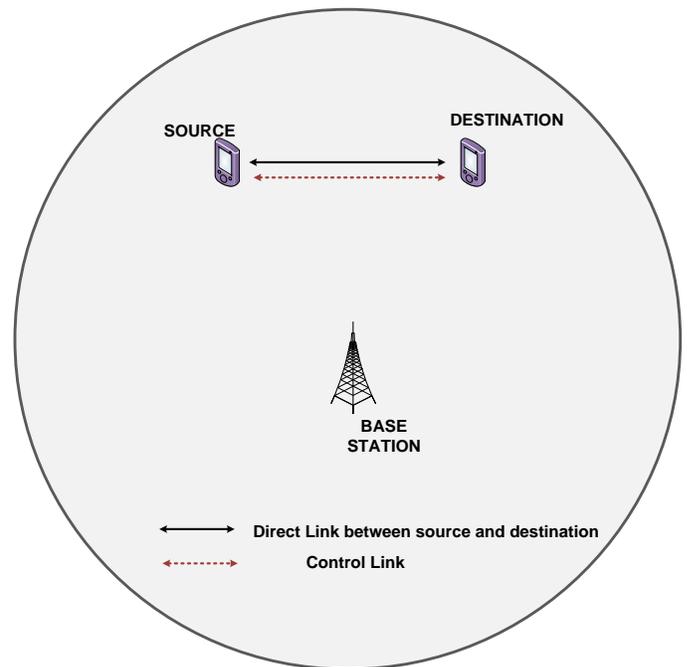

Fig.7. Direct communication between devices (Direct D2D) with device controlled link establishment.

The two-tier cellular network architecture is advantageous over the conventional cellular architecture. The benefits offered are as follows:

*1. One hop communication:* The devices can communicate with each other through a single hop. Lesser resources are then required for the communication, resulting in an efficient utilization of the spectrum. Since proximity users directly communicate with each other in D2D communication, latency is greatly reduced. These are desirable aspects in a cellular network. The mobile network operators are also benefitted by these aspects of D2D communication.

*2. Spectrum Reusability:* With D2D communication in cellular networks, same spectrum is shared by the D2D users as well as the cellular users. This supports spectrum reusability, thereby improving the spectrum reuse ratio.

*3. Optimization of Power Levels:* Since D2D links exist between proximate devices, over a small distance, transmission power is less. This enhances the battery life of the devices. As a result, higher energy efficiency can be achieved with D2D communication in cellular networks.

*4. Improved Coverage Area:* As discussed in (1) and (3), D2D communication is possible with relays. This supports communication over greater ranges, thus increasing the overall coverage area.

Optimal density of D2D users in a network is demonstrated in [14]. In spite of the number of advantages that are offered by D2D communication over the conventional cellular communication, some limitations exist. The authors in [15] discuss about possibility of use of D2D communication within the cellular systems. Feasibility of D2D communication is determined by the distance restriction. Another concern is the interference, which may be between the users of the same tier or different tiers. In cases of base-station assisted D2D communication ((1) and (2)), the BS essentially acts as a central controlling entity and can overcome interference problem to some extent. The base station (BS) manages spectrum allocation and aids in avoiding interference among the devices. In device-assisted D2D communication ((3) and (4)), there is no central controlling entity. These communication techniques are more challenging than the other two. For optimum performance of D2D communication in the cellular networks, smart interference management schemes, supporting optimal resource allocation need to be designed. A considerable amount of literature is available in this context, and offers a wide range of opportunities to the researchers to further explore these areas.

Prior to direct transmission of information between the devices, these need to find each other. Device discovery can be possible by a periodic broadcasting of the device identity. Distance constraint is generally considered, for D2D pair formation. Peer discovery and mode selection is an open research issue in device-to-device (D2D) communication. For any cellular network, a major concern is security. When exchanging information through relays, as in (3), network security must be assured. This can be made possible by 'closed access', where a list of trusted devices is prepared by every device belonging to the device tier. If a device, under the relay scenario, does not find some devices in its own list, it communicates in the macro cell tier. The issue of security, with machine-to-machine (M2M) communication taken as reference, is discussed in [16-18], [19], [20]. The base station has the capability to authenticate the devices that are acting as relays, and use encryption to maintain privacy for the information of devices.

## IV.  INTEGRANT FEATURES OF D2D

Originally, the concept of device-to-device (D2D) communication was used for sensor networks, ad hoc networks and mesh networks. The devices communicated in a distributive fashion, in the industrial, scientific, medical (ISM) band, in the absences of any controlling entity. Nowadays, however, in LTE-A and the next generation networks (NGNs),

D2D communication is gaining popularity for use in the licensed band. Formation of direct links is useful for the improvement in the overall network performance, and also to the devices in terms of energy efficiency and complexity. A number of features of 5G networks can be integrated with device-to-device (D2D) communication (Fig.8.), which acts as an enabler for D2D communication in the existing cellular networks. Some of these have been briefly listed below.

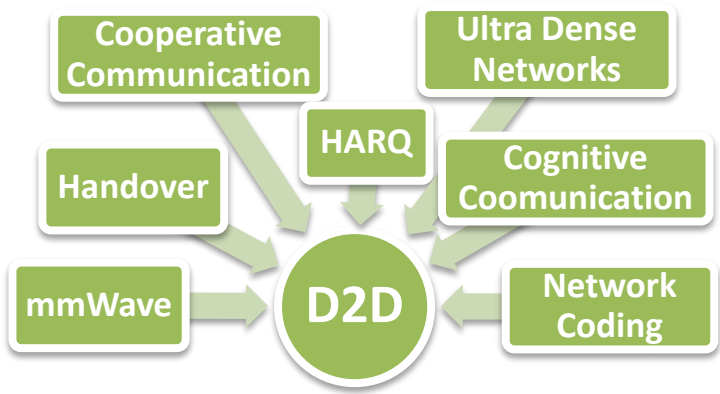

Fig .8 . D2D Integrant Features

### A. Millimeter wave D2D Communication

A promising technology of the future 5G networks is millimeter wave (mmWave) communication, providing multi-gigabits-per-second to the user equipments (UEs). It operates over a wide frequency band of 30GHz to 300GHz. Efficient utilization of the bandwidth is feasible by enabling device-to-device (D2D) communication in the next generation networks. Using D2D communications in mmWave cellular networks, a number of direct concurrent links can be supported, resulting in an improved network capacity. Also, simultaneous connections can be supported in mmWave networks due to the highly directional antennas and high propagation loss in mmWave communication. A scheduling mechanism for downloading of popular content in mmWave small cells, exploiting D2D transmissions has been proposed in [21], resulting in an overall improvement in transmission efficiency. As per the simulation results, the proposed scheme results in reduced latency and enhanced throughput. This clearly reveals the benefits of D2D communication at mmWave frequencies.

In mmWave 5G cellular networks, two types of D2D communications are possible - local D2D communication and global D2D communication. In local D2D communications, if the LOS path is blocked, then a path is developed between the two devices associated with the same base station, with the help of relays or directly. In global D2D communications, devices associated with different BSs are connected through the backbone networks, via hopping.    But, D2D connections in mmWave networks can suffer interference [22]. This is possible in case there are multiple D2D communications within the cellular network (local D2D communications).

Coexistence of local and global D2D communication in the network results in interference between local D2D communications and between B2B/D2B (Base station to base station/ Device to Base station) communications. Due to the highly directional nature of mmWave communication, high data rate B2B communications are supported in the cellular networks. Since mmWave communication use directional antennas, resource sharing schemes must take into account the directional interference as well for such scenarios, for efficient spectrum utilization. Although use of directional antennas is advantageous in terms of enhanced network capacity and spatial reuse, there are certain challenges also associated with it. Generally, problems arise in case of neighbor discovery, like deafness problem, and tend to promote research in this field. The problem related to blockage and directionality in mmWave communication has been solved using cross-layer modeling and design techniques [23]. A major problem for mmWave propagation is unavailability of a standard channel model.

### B. Cooperative D2D Communication

Cooperative communication is a focal technology in the cellular networks today. For D2D communication, their impact is expected to be remarkable. When the D2D pairs are far away from each other, the direct link between the users is not good enough for communication [24]. Here is where cooperation plays an essential role. Cooperation aids in improving the quality of D2D communication for data offloading between the UEs. It enables interference reduction as well as increased network coverage. In order to depict cooperative networking, a scenario of cooperative communication has been shown in Fig.9.

In case of cooperative D2D communication, the network adaptively decides the communication mode as underlay, overlay or cooperative relay mode on the basis of channel quality and data rate requirement.  Selection of relay is an important issue in cooperative D2D networks. Since a large number of relays can be used, the relay selection needs to be optimum and efficient. Relay selection methods have been proposed in [25]. These algorithms help the BS to choose the best of all the relays. In case of relay selection, generally the BS is assumed to play a passive role because using centralized methods in the selection process increases load on the BS increases. Selecting relays in a distributive manner eliminates relays that are not proper.

 Cooperation is based on social reciprocity, and trust, is discussed in [26]. The authors have evaluated an efficient D2D cooperation strategy by proposing a game-theoretic approach. A cooperative multi-hop D2D scenario is discussed in [27], which results in boosting of data rate. Though cooperation contributes towards improving system performance and QoS, but a large amount of UE power is also consumed. This needs to be optimized. In literature, cooperation among D2D users and also between D2D users and cellular users is widely studied.

### C. Handover in Device-to-Device Communication

When devices are undergoing D2D communication, they enter into the neighbor cells at some point or the other. When the two UEs are in close proximity, they undergo a joint handover. Under certain circumstances, the devices may not be in proximity or one of them may get handed over to some neighboring cell, resulting in a half handover. Very less literature is available on handover of D2D communication.

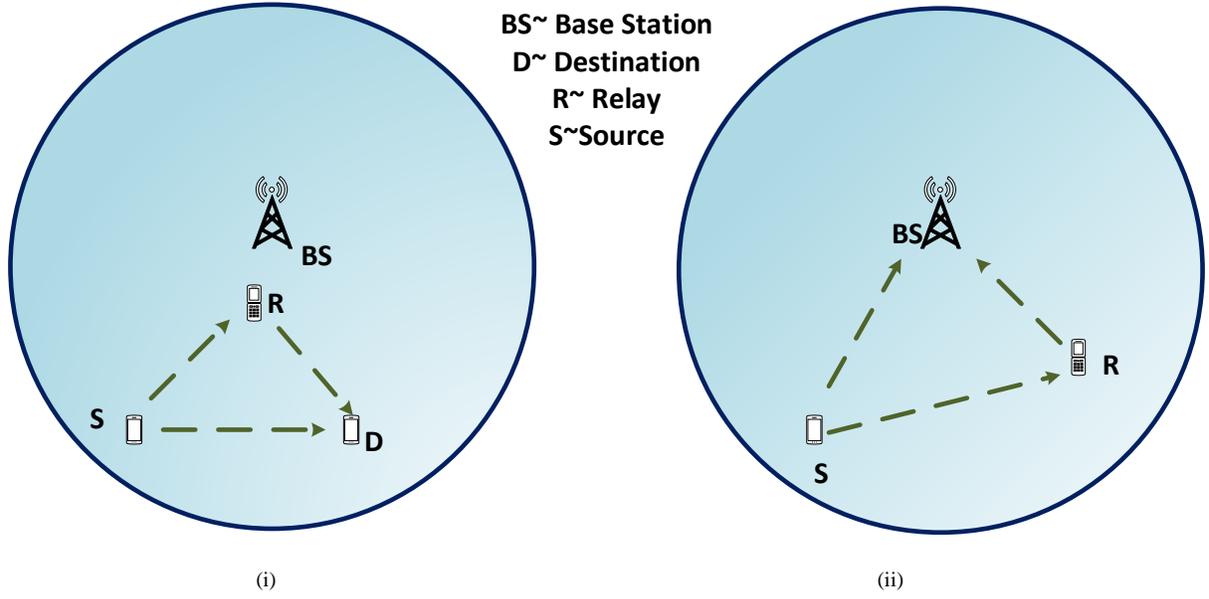

(i)                      (ii)

Fig. 9. Cooperative D2D communication; R serves as Relay for D2D communication in both the scenarios, supporting cooperation

A basic and effective handover algorithm is the handover decision method [28], involving the use of a number of variables referred to as, Handover Margin (HOM), Time to Trigger (TTT) timer, LTE threshold ($LTE_{th}$), D2D threshold ($D2D_{th}$) and Time to Trigger of D2D ($TTT_D$).

HOM is a constant variable representing a threshold of the difference between the strength of the received signal to the source eNB and the strength of the received signal to the target eNBs. The strength of the received signal is called reference signal receiving power (RSRP), in an LTE system. It is ensured by the value of HOM that the target eNB is the most appropriate for Proximity Services (Prose). Value of TTT is the time interval required to satisfy HOM condition, as stated in [28].Once TTT condition is satisfied, then the handover action can be successfully completed. Different values of TTT can be used by ProSe UEs. Unnecessary handovers, called "Ping-Pong effect" can be reduced by HOM and TTT. LTE Threshold ($LTE_{th}$) is a constant variable which represents whether the basic services can be provided by the source eNB to the ProSe UEs or not. D2D threshold ($D2D_{th}$) is used to check the radio signal strength of D2D quality. The conditions for triggering handover are

$$RSRP_T > RSRP_S + HOM \quad (1)$$

$$HOTrigger > TTT \quad (2)$$

Here, $RSRP_T$ and $RSRP_S$ are the values of RSRP from target and source eNBs, respectively. THE $HOTrigger$ is the handover trigger timer which turns on as soon as (1) is satisfied. The handover decision is made by the eNB provided all requisite conditions are satisfied. On the basis of the D2D handover decision method, a joint or a half handover procedure can be selected, or even no handover. In case of joint handover, a collective handover of all the devices takes place to the target eNB, while in case of half handover; one of the UEs is handed over to the target eNB while the other remains connected to the source eNB. When handover occurs, there is exchange of some unnecessary control overhead as well, between the devices. A general handover scenario has been depicted in Fig.10, representing handover of UE1 from one base station to another. Mobility management solutions have been provided in [29] where two schemes for smart mobility management have been proposed: D2D-aware handover and D2D-triggered handover. The simulation results illustrate that these schemes reduce end-to-end latency in the D2D communication and reduce signaling overhead as well, within the network. Vertical and horizontal handover are efficient for reducing energy consumption in heterogeneous networks [30].

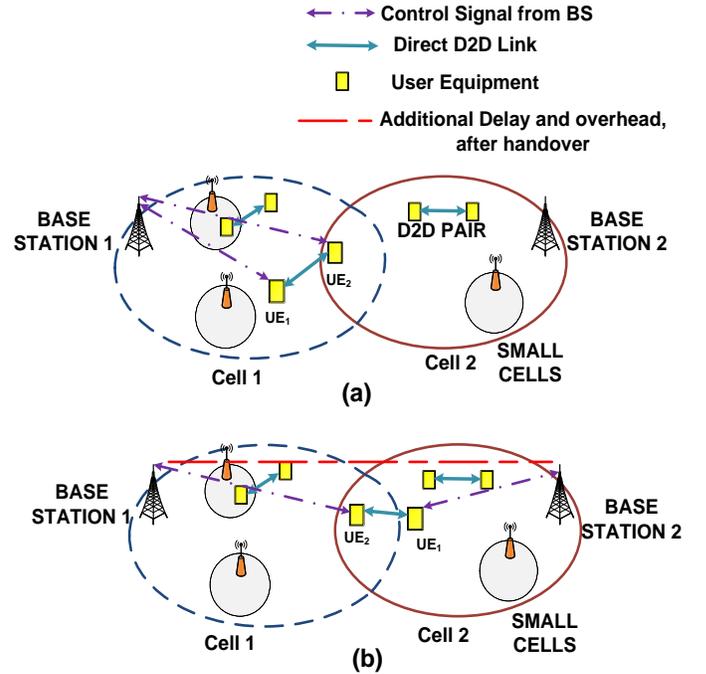

Fig.10. Regular Handover Scenario, (a) Before Handover; (b) After Handover from cell1 to cell2

### D. Hybrid Automatic Repeat Request (HARQ) Operation

Automatic Repeat Request (ARQ) retransmission and forward error correction are combined in HARQ. It tends to make D2D communication more robust. In D2D communication, two types of HARQ exist- Direct and Indirect [31]. In case of indirect HARQ, an ACK/NACK is sent by the D2D receiver to the eNB which is then further relayed to the D2D transmitter. Reusing of uplink and downlink channels is possible with indirect HARQ. The D2D receiver directly sends an ACK/NACK to the D2D transmitter, in case of direct HARQ. It can be used either in in-coverage or an out of coverage scenario.

A cellular HARQ phenomenon has been depicted in Fig.11. The figure shows multicasting of packets by the BS to the UEs in the network. The HARQ feedback message provides the receiving status of the packets, at the UEs. Depending on whether a packet is received or not, an acknowledgement/negative acknowledgement (ACK/NACK) is sent by the UEs. In case the BS receives a NACK, it retransmits the packet. This technique consumes a large amount of energy and involves significant signaling overhead. A compressed HARQ mechanism has been proposed in [32], in a network with underlay D2D communication, which provides better results in terms of signaling overhead. This mechanism is highly efficient for multicast services and performs better than conventional D2D multicast. The authors of [33] propose a cross layer design based on HARQ. Three types of HARQ have been discussed in this design: Type I HARQ, Type II HARQ and Type III HARQ. Using HARQ and cross layer optimization effectively results in improving the D2D transmission rate, and throughput. Thus, incorporation of HARQ in D2D communication will result in an efficient error correction within the network.

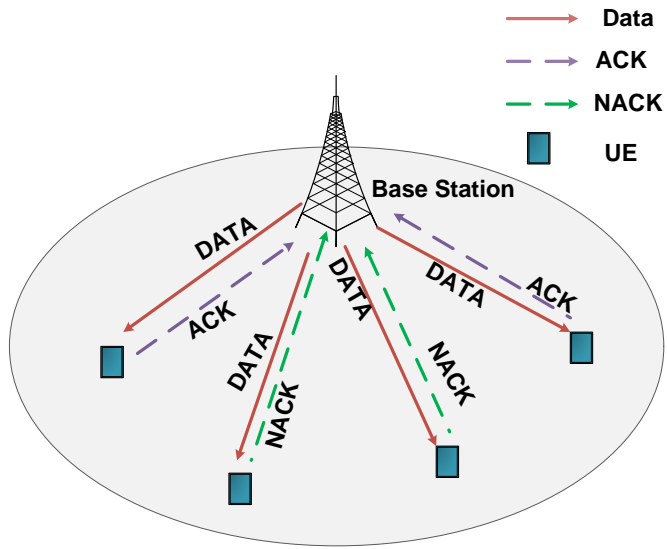

Fig.11. HARQ process

### E. D2D Ultra Dense Networks

An important concern of operators today is offloading cellular data. With the growth in the use of smart phones and tablets, the core and access networks tend to overload. Traffic offloading is necessary in such a scenario, so as to free up the loaded path by providing alternate paths to the traffic. In 3GPP system architecture (Release 10), two offloading solutions have been provided, which are LIPA and SIPTO [34]. Another offloading technique from 3GPP is device-to-device (D2D) communication. D2D offloading capability is more advantageous in comparison to LIPA and SIPTO, as D2D offloading avoids radio congestion as well, apart from offloading the core network. This results in an enhancement in network capacity. The offloading solutions from 3GPP have been compared in Table III. A detail of these techniques is provided in [35].

Apart from these techniques, small cells provide an efficient means for offloading the traffic, and aid the other offloading techniques as well. With cell size getting smaller over the generations, there is less competition among the users for resources, yielding a substantial increase in spectrum efficiency. Small cells include picocells, microcells and femtocells, varying in the cell sizes and transmission power. Deployment of a large number of low power small cell base station (SBSs) results in ultradense networks (UDNs). Such a deployment helps in frequency reuse and controls interference. Ultra Dense networks (UDNs) have more number of nodes (UEs) per unit area. UDN has recently been accepted as an important enabling technology for enhancing the network capacity.

TABLE III
COMPARISON OF 3GPP OFFLOADING SOLUTIONS

| S No | Feature | SIPTO | LIPA | D2D Communication |
|---|---|---|---|---|
| 1. | Definition | Offloads selected IP traffic to the internet locally, as well as at macrocellular access networks | Allows offloading traffic directly to a local network, which is connected to the same H(eNB), as the UE | Offloads traffic at the radio access network, as well as the core network |
| 2. | Qos | Not maintained | Not maintained | Maintained |
| 3. | Offload Points | At or above eNBs | At or above eNBs | Data Offload points positioned at mobile terminals |
| 4. | 3GPP Release | Rel 10 | Rel 10 | Rel 12 |

D2D along with small cells [36], both play a key role in offloading traffic from the eNB. D2D mainly focuses on offloading proximity services while hot-spot traffic is offloaded by the small cells. Integration of these two technologies results in Ultra-dense 5G deployments, as shown in Fig.12. UDNs as an important component of the next generation networks (NGNs) have been discussed in [37]. It is expected to enable higher data rates and lower delays within the network. Working of UDN in the mmWave band will result in a contiguous bandwidth of about 2GHz. The system level performance of UDN has been evaluated in [38]. The simulation results show increase in QoS with increasing number of SBSs. However, there are very high chances of interference between the macro-cell links, the D2D links and the small cell links. This problem can further worsen if the D2D links are from different cells. Also, deployment of SBSs is a big challenge. All these aspects need to be critically addressed.

of the network design parameters. A CR-assisted D2D communication in cellular networks has been investigated in [41], in which, the UEs access the spectrum by a mixed underlay/overlay sharing of spectrum. Cognitive and energy harvesting-based D2D communication has been modeled in [42]. Its proposed model is evaluated on the basis of the stochastic geometry, which shows that the overall QoS of the cellular network improves with cognitive D2D communication, when network parameters are tuned carefully. The use of D2D communication for vehicular communication is discussed in [43], with the use of cognitive radio for offloading vehicular traffic. The results show reduction in transmission delay. A combination of D2D and cognitive communication makes D2D communication very diverse [44].

*G. Network Coding*

A potential technique for the overall throughput improvement of a network is network coding. The transmitting nodes, with network coding, tend to combine the packets before transmission. This reduces the amount of routing information. Network coding in D2D communication helps in reducing power consumption, interference, etc [45]. It also provides security and communication efficiency. This has been discussed in [46], using the protocols: CORE and PlayNCool. Due to its unique advantages, network coding enables throughput improvement, delay reduction and energy efficiency in the D2D communication. Though there are number of advantages of network coding, but it requires a large amount of resources (both time and radio) for decoding the data received at the D2D node. Additionally, since a number of packets are combined, uniqueness of the coefficients cannot be guaranteed. As a result, this remains an open research field for the researchers.

The above mentioned are a few features which can be used in conjunction with D2D communication. Once successfully implemented, these will result in numerous advantages to the service providers, as well as the subscribers. The overall utility of the cellular networks will be greatly enhanced, as presented in the preceding discussion. Issues related to the above mentioned features of wireless networks require attention and further research.

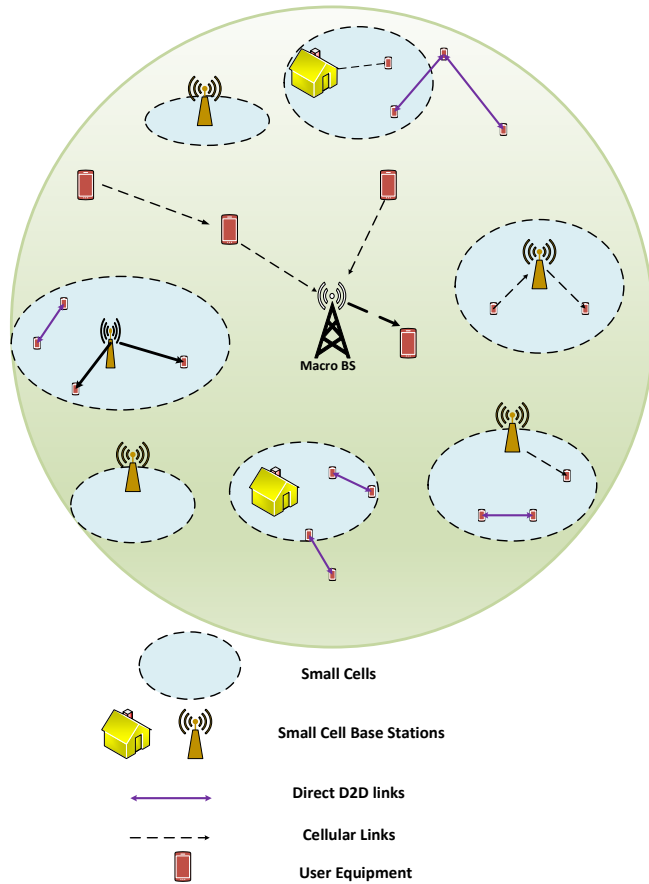

Fig.12. Ultra dense networks with D2D communication

*F. Cognitive D2D*

Cognitive communication has played an essential role in improvement of spectrum efficiency by enabling the use of vacant bands by secondary users without causing any hindrance to the primary users. Cognitive Radio Networks (CRNs) offer a class of networks that have the ability to change their operating parameters, on the basis of their interaction with the surrounding environment. Unused part of the spectrum can be utilized by CRNs by spectrum sensing. Consensus based algorithms for cooperative spectrum sensing has been given in [39].

The use of cognitive D2D reduces the burden of frequency allocation on the operator. Additionally, sensing and reusing of ISM band resources is possible with cognitive D2D. The ongoing communication within the ISM band remains unaffected. In inband D2D communication (discussed in next section), cognitive spectrum access (CSA) can result in efficient resource utilization and interference management. Two spectrum access techniques, D2D-unaware spectrum access and D2D-aware spectrum access, have been discussed in [40]. The CSA scheme can be optimized by finest selection

## V. KEY OPEN CHALLENGES IN D2D

Device to device (D2D) communication may use the licensed spectrum (in band) or the unlicensed spectrum (out band) for direct link formation [47]. Inband D2D communication is categorized as underlay and overlay. Underlay D2D communication allows set up of direct links and cellular links in the cellular spectrum. In overlay D2D, on the other hand, a dedicated portion of the available spectrum is used for Device-to-Device (D2D) communication, with rest of the spectrum used for cellular communication. As out band D2D communication exploits the unlicensed spectrum for the formation of direct links, it is categorized as autonomous and controlled. When controlled, the radio interfaces in D2D are managed by the eNB, while in autonomous, these are coordinated by the user equipments (UEs) themselves. Interference between D2D users and cellular users is no issue in out band D2D, but coordination of the communication in the unlicensed band requires a second radio interface (like, Wi-Fi Direct [48], Bluetooth [49], ZigBee [50]). The categorization of D2D communication has been depicted in Fig.13. To utilize the limited available spectrum in the most efficient manner, one must know where to use which category of D2D communication.

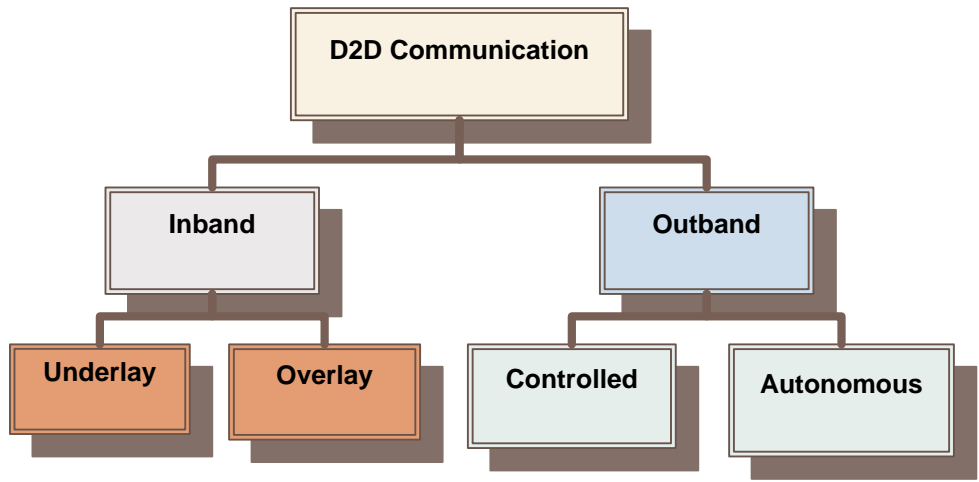

Fig.13. Types of D2D Communication [47]

For implementing D2D communication in cellular networks, a number of key issues need to be addressed. To obtain complete advantage of Device to Device (D2D) communication, overcoming these issues efficiently, is important. Some of these are listed below, and available literature considers in band as well as out band D2D.

### A. PEER DISCOVERY

Since D2D communication is gaining popularity, identifying efficient means of discovering proximate users has become necessary. The process of peer discovery should be efficient, so that D2D links are discovered and established quickly. It is also important for ensuring optimum throughput, efficiency and resource allocation within the system. Setting up of direct links requires devices to discover each other first. Once discovered, direct links are set up, and then occurs transmission over those links. Researchers are working on different approaches for device discovery. In [51], spatial correlation of wireless channels is considered for low power peer discovery. The simulation results show that peers can be discovered with very low power consumption. It provides a very accurate method of peer discovery. Peer discovery techniques can be restricted discovery and open discovery [52]. In case of restricted discovery, the UEs cannot be detected without their prior explicit permission. This thus maintains user privacy. In case of open discovery, UEs can be detected during the duration for which they lie in proximity of other UEs. From the perspective of the network, device discovery can be controlled by the base-station either tightly or lightly [53], [54].

The authors in [55] propose a neighbor discovery technique which is based on power vectors, and considering a time-variant channel. It is a low complexity algorithm, where the probability of a false detection is close to zero. Energy required to support D2D communication is high. For an energy efficient network, a device discovery technique is proposed in [53]. A social-aware peer discovery scheme has been proposed in [56]. The scheme enhances the network performance by improving the data delivery ratio, exploiting the social information only. An effective network-assisted technique for device discovery has been proposed in [57] for the support of device-to-device communication in LTE networks. The results show that the probability of device discovery is quite high in this technique, for a certain discovery interval.

TABLE IV
METHODS OF PEER DISCOVERY

| REFERENCE NO. | METHOD OF DISCOVERY |
|---|---|
| [51] | Low power Discovery |
| [52] | • Restricted Discovery<br>• Open Discovery |
| [53] | Energy efficient device discovery |
| [55] | Discovery based on power vectors |
| [56] | Social aware peer discovery |
| [57] | Network-assisted discovery |
| [58] | Sound Referencing Signal for neighbor discovery |
| [59] | • Bluetooth Discovery<br>• Wi-Fi Device Discovery<br>• Wi-Fi Direct Device Discovery<br>• IrDA Device Discovery<br>• Network Assisted Discovery<br>• Packet and Signature-based Discovery<br>• Request Based Discovery<br>• Direct Discovery |

In [58], the authors propose neighbor discovery with the use of a sounding reference signal (SRS) channel. The uplink transmissions of cellular users play an essential role in finding the neighbors. Neighbor discovery under unknown channel statistics is also considered. A review to various techniques for device discovery is given in [59]. These include Bluetooth discovery, Wi-Fi (Ad Hoc) Device Discovery, IrDA Device Discovery. Request based discovery, Direct Discovery, Request based Discovery, Packet and Signature-based

Discovery and Network-Assisted discovery. A summary of the various peer discovery methods is given in Table IV.

On completion of peer discovery, session setup takes place. For the setting up of sessions, two methods have been developed; IP based detection and dedicated D2D signaling. The existing literature mainly focuses on single cell scenarios, for device discovery and session setup. Works on multi cell scenario is more beneficial as it supports efficient resource utilization. Device-to-Device (D2D) discovery and session set up is a very challenging job, since it needs cooperation from the adjacent base-stations (BS).

*B. RESOURCE ALLOCATION*

After device discovery, availability of resources is important for enabling communication over the direct links. Radio resource allocation is thus important for enhancing the spectral efficiency of D2D communication, underlaying cellular communication. Resource allocation strategies in D2D communication can be centralized or distributed. Centralized techniques [19] cause complexity in case of large networks while distributed techniques [52] tend to decrease the device complexity. The distributed techniques improve the scalability of the D2D links. Hybrid solutions also can be provided and are an area of research. A number of different techniques are available under the literature survey. For obtaining maximum throughput, D2D communication can operate in a number of modes. These can be:

*Silent Mode*: In this mode, the D2D devices stay silent and cannot transmit because of lack of resources. Spectrum reuse, as a result, is not possible.

*Dedicated Mode*: In this mode, some of the available resources are dedicated for the D2D users, to be used for direct transmission.

*Reuse Mode:* In this mode, uplink or downlink resources of the cellular users are reused by the D2D users.

*Cellular Mode*: In this, conventional communication occurs, through the eNBs and D2D data is transmitted.

An improvement in the spectrum efficiency can be achieved by the use of reuse mode. Interference management is better with the dedicated and cellular modes. However, these two modes maybe inefficient to maximize the overall network throughput. The decision for resource sharing is made by the base station. When the D2D links and cellular links reuse the same resources, it is referred to as non-orthogonal sharing, and when they do not share the same resources, it is referred to as orthogonal sharing. Better resource utilization efficiency is achieved by non-orthogonal sharing.

Considering the spectrum efficiency of LTE-A networks, a resource allocation strategy is proposed in [60], for minimization of the transmission length of D2D links. An NP-complete [61] problem is formulated as a Mixed Integer Programming. A low complexity column generation method solves the resource allocation problem in D2D communication. Another technique for resource allocation is provided in [62], maximizing throughput of the network. The cellular services are given the higher priority over the D2D communication. For evaluating a single-cell scenario, a system with D2D communication underlaying cellular communication is considered. Resource sharing as orthogonal sharing, non-orthogonal sharing and cellular operation is discussed. In [63], optimal resource utilization is achieved by cluster partitioning. In [64], a method is proposed for overall throughput improvement of the system, along with enhancement in spectral efficiency through power allocation and admission control. A Heuristic Location Dependent Resource Allocation Algorithm has been proposed in [65]. It is customized to vehicle-to-vehicle (V2V) communication, aiming to give prime priority to safety of V2V communication. Resource pooling has been proposed in [53].

A semi-persistent resource sharing algorithm has been proposed in [66], in which inter-cell and intra-cell scenarios have been considered. This algorithm improves the overall throughput of the network. Non-orthogonal resource sharing is discussed in [67], [68], considering the maximum transmit power constraint. The authors of [69] propose another optimal resource allocation technique that is able to significantly improve the sum throughput of D2D as well as cellular communication in a network. In order to improve the overall network throughput and user satisfaction ratio, the authors of [70] introduce a time-division scheduling (TDS) algorithm for efficient utilization and allocation of resources, using non-orthogonal sharing mode. Based on the improved proportional fairness algorithm, the authors in [71] propose adaptive time division scheduling algorithm, in which D2D pairs are adaptively allocated to the timeslots, unlike [70]. A brief overview of some of the mentioned algorithms has been given in Table V.

On the basis of various algorithms discussed so far, we have observed that compared to other well known schemes, [64], [71] and [70] have provided the best performance. The D2D access rate, throughput gain, fairness and user satisfaction ratio have been maximized in these algorithms, which is desirable from the user perspective as well as the service provider.

*1) Network Model*

A single cell scenario, with the base-station (BS) at the centre, a D2D pair and cellular users is considered, as shown in Fig. 14, with D2D communication underlaying cellular communication. The users that are capable of carrying out direct D2D communication are identified by the base station. The location information of all users and the channel state information (CSI) are provided to the base station (BS) through the global positioning system (GPS) receiver available on the user equipments (UEs). There are high chances of potential interference among the users, as depicted by the interfering signals, in Fig. 14. A D2D link exists between and the D2D transmitter ($D_{tx}$) and D2D receiver ($D_{rx}$,), in accordance with the distance constraint, $D \leq d_0$, where D is the distance between $D_{tx}$ and $D_{rx}$ and $d_0$ is the maximum distance for direct communication.

As an assumption, it is considered that each cellular user is allotted equal number of resource blocks (RBs). The RB which is allocated to a particular user is shared by a single D2D pair, so as to avoid interference among the D2D pairs. A single pair can share resources of multiple cellular users in the network. The network is assumed to contain *m* number of cellular users, *n* number of D2D pairs and *k* number of resource blocks. Let the channel gain between base station and a cellular user be given by $g_{bcu(i)}$, channel gain between D2D

13TABLE V
RESOURCE ALLOCATION ALGORITHMS FOR DEVICE-TO-DEVICE COMMUNICATION

| REFERENCE NO. | *ALGORITHM* | *DESCRIPTION* | *OBJECTIVE* |
|---|---|---|---|
| [53] | Resource Pooling | Allows resource reuse between D2D and cellular users, taking advantage of hop gain | To achieve increased throughput, power saving and higher spectrum efficiency |
| [61] | Column generation method | A heuristic algorithm that detects maximum number of active D2D links which are capable of transmitting simultaneously in every time slot, satisfying the access pattern constraints | With increasing power consumption, reduction in transmission length of D2D connections for dense networks |
| [62] | Resource sharing in traditional cellular and direct D2D communication | Transmissions in orthogonal, non-orthogonal and cellular resource sharing modes are optimized in order to maximize the overall sum rate | Optimization of sum rate, taking into consideration resource allocation and power control, and also adhering to transmit power/energy constraints |
| [63] | Cluster partitioning and relay selection | An intra-cluster D2D retransmission scheme in which cooperative relays are adaptively selected through multicast retransmissions | To achieve optimal resource utilization |
| [64] | Admission control and power allocation | A maximum weight bipartite technique is used to find suitable D2D pair for each cellular user; optimum power is allocated to D2D pairs and their cellular partners | Improvement in spectral efficiency, with enhanced system throughput |
| [65] | Heuristic Location Dependent Resource Allocation algorithm | Persistent resource allocation applied to the network considered, with fixed reservation of resources | Feasible for QoS controlled and services demanding strict reliability. This algorithm aims at reduction in the signaling overhead and interference of the network under consideration |
| [66] | Uplink Semi-Persistent Scheduling Resource Reuse Algorithm | The D2D users reuse the UL semi persistent resources for minimum interference. The algorithm takes into consideration inter cell as well as intra cell interference between D2D and cellular links | Improvement in overall system throughput and reduction in the interference among D2D links |
| [70] | Time Division Scheduling (TDS) Algorithm | The entire scheduling period of the base station is divided into $n$ equal number of time slots. A location dispersion principle is used to allocate the D2D pairs in a balanced number, to the slots | A significant improvement in system throughput, along with high D2D user satisfaction ratio |
| [71] | Adaptive Time Division Scheduling Algorithm | D2D pairs are adaptively allocated to a series of timeslots, using improved proportional fairness algorithm; resources of cellular users are allocated to the D2D pairs assigned to the timeslots | High system throughput and better fairness |

transmitter and receiver $g_{d(j)}$, the gain of interference link from base station to $D_{rx}$ be denoted by $g_{bd(j)}$, and gain of interference link between $D_{tx}$ to cellular user be $g_{d(j)cu(i)}$. These channel gains tend to contain distant dependent path loss as well as shadowing path loss. The base station is assumed to allocate resource blocks, for efficient utilization of cellular resources.

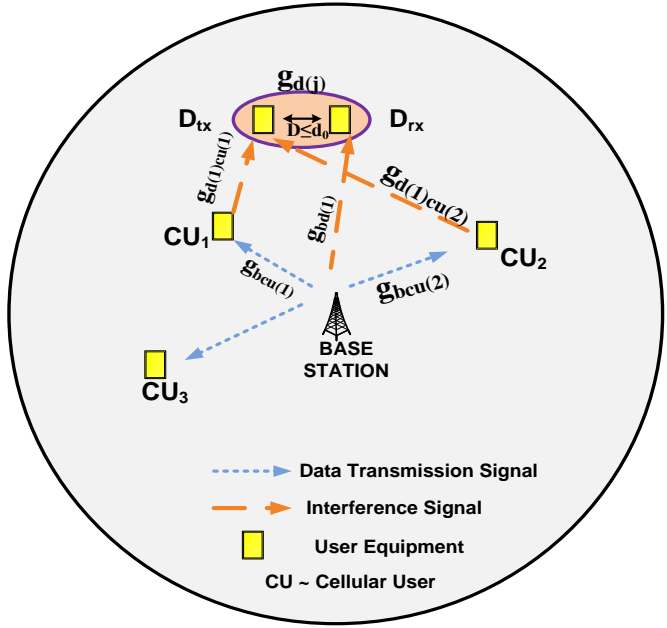

Fig. 14. Network Model

The transmission power of the BS and $D_{tx}$ are given by $P_{BS}$ and $P_{DD}$, respectively

When $j^{th}$ D2D pair, shares $k^{th}$ RB with $i^{th}$ cellular user, $j \in n$ and $i \in m$, the SINRs at the cellular user and $D_{rx}$ can be represented by

$$\mathcal{S}_{cu(i)d(j)} = \frac{P_{BS}\, g_{bcu(j)}}{\acute{\eta} + P_{DD}\, g_{d(j)cu(i)}} \tag{3}$$

$$\mathcal{S}_{d(j)} = \frac{P_{DD}\, g_{d(j)}}{\acute{\eta} + P_{BS}\, g_{bd(j)}} \tag{4}$$

On the other hand, when a cellular user does not share its RB with any pair, then the its SINR is given by

$$\mathcal{S}_{cu(i)} = \frac{P_{BS}\, g_{bcu(i)}}{\acute{\eta}} \tag{5}$$

When a cellular user shares RB, the data rate of $i^{th}$ cellular user after and before sharing of resources is given respectively by,

$$R_{cu(i)d(j)} = BW * \log(\mathcal{S}_{cu(i)d(j)}) \tag{6}$$
$$R_{cu(i)} = BW * \log(\mathcal{S}_{cu(i)}) \tag{7}$$

where BW denotes the bandwidth of a single RB.

When a D2D pair shares a RB with a cellular user, the channel rate is given by

$$R_{d(j)} = BW * \log(1 + \mathcal{S}_{d(j)}) \tag{8}$$

When $i^{th}$ cellular user shares $k^{th}$ resource block with $j^{th}$ D2D pair, it suffers interference from $D_{tx}$, resulting in a decrease in transmission rate. This decrement in the data rate is represented by

$$\Delta R_{d(j)} = R_{cu(i)} - R_{cu(i)d(j)} \tag{9}$$

There is an increment in the overall system throughput, after the sharing of resources, since rate of the D2D pair increases much more, compared to the decrement in the cellular user rate. This throughput increment is represented by

$$\Delta T_{cu(i)d(j)} = R_{d(j)} + R_{cu(i)d(j)} - R_{cu(i)}$$
$$= R_{d(j)} - \Delta R_{d(j)} \tag{10}$$

The optimization problem is formulated as

$$\Delta T_m = \arg\max \sum_{k=1}^{l} \sum_{j=1}^{n} W_{k,j}\, (R_{d(j)} + R_{cu(i)d(j)} - R_{cu(i)}) \tag{11}$$

subject to

$$\mathcal{S}_{cuiDj} \geq SINR_{cu}^{t} \tag{12}$$

$$\mathcal{S}_{Dj} \geq SINR_{D}^{t} \tag{13}$$

(12) and (13) are defining the threshold SINR values for cellular user and D2D users, respectively. Thus, throughput maximization is accomplished by the above analysis, in cellular networks with underlaying D2D communication. The condition for minimum number of RBs required for the $j^{th}$ D2D pair to maximize throughput is stated as

$$\begin{cases} (N_{min(j)} - 1)\, R_{d(j)} < k\, R_{min} \\ N_{min(j)}\, R_{d(j)} \geq k\, R_{min} \end{cases} \tag{14}$$

Where is $R_{min}$ is the minimum data rate required by the D2D users, for optimal resource sharing between the users.

The various symbols used in the network model have been summarized in Table VI.

TABLE VI
SYMBOLS USED IN NETWORK MODEL

| Symbol | Meaning |
|---|---|
| $g_{bcu(i)}$ | Channel gain between BS and CU |
| $g_{d(j)}$ | Channel gain between D2D pair |
| $g_{bd(j)}$ | Channel gain of interference link between BS and $D2D_{rx}$ |
| $g_{d(j)cu(i)}$ | Channel gain of interference link between BS and $D2D_{tx}$ |
| $\mathcal{S}_{cu(i)d(j)}$ | SINR at $i^{th}$ CU on sharing RB with $j^{th}$ D2D pair |
| $\mathcal{S}_{d(j)}$ | SINR at $j^{th}$ D2D pair on sharing RB with $i^{th}$ CU |
| $\mathcal{S}_{cu(i)}$ | SINR of $i^{th}$ CU not sharing any RB with D2D pair |
| $R_{cu(i)d(j)}$ | Data rate of $i^{th}$ CU after sharing RB with $j^{th}$ D2D pair |
| $R_{cu(i)}$ | Data rate of $i^{th}$ CU before sharing RB |
| $R_{d(j)}$ | Data rate of $j^{th}$ D2D pair |
| $\Delta T_{cu(i)d(j)}$ | Throughput increment on RB sharing |
| $\acute{\eta}$ | Noise Power Density |
| $N_{min(j)}$ | Minimum number of RBs required for the $j^{th}$ D2D pair for maximizing throughput. |
| $R_{min}$ | Minimum data rate required by the D2D users |

For the next generation networks, in order to meet the rising demands and requirements of the mobile network operators, architecture has been proposed (as depicted in Fig.15.). Essential network requirements are expected to be met efficiently, through this proposed architecture for resource allocation.

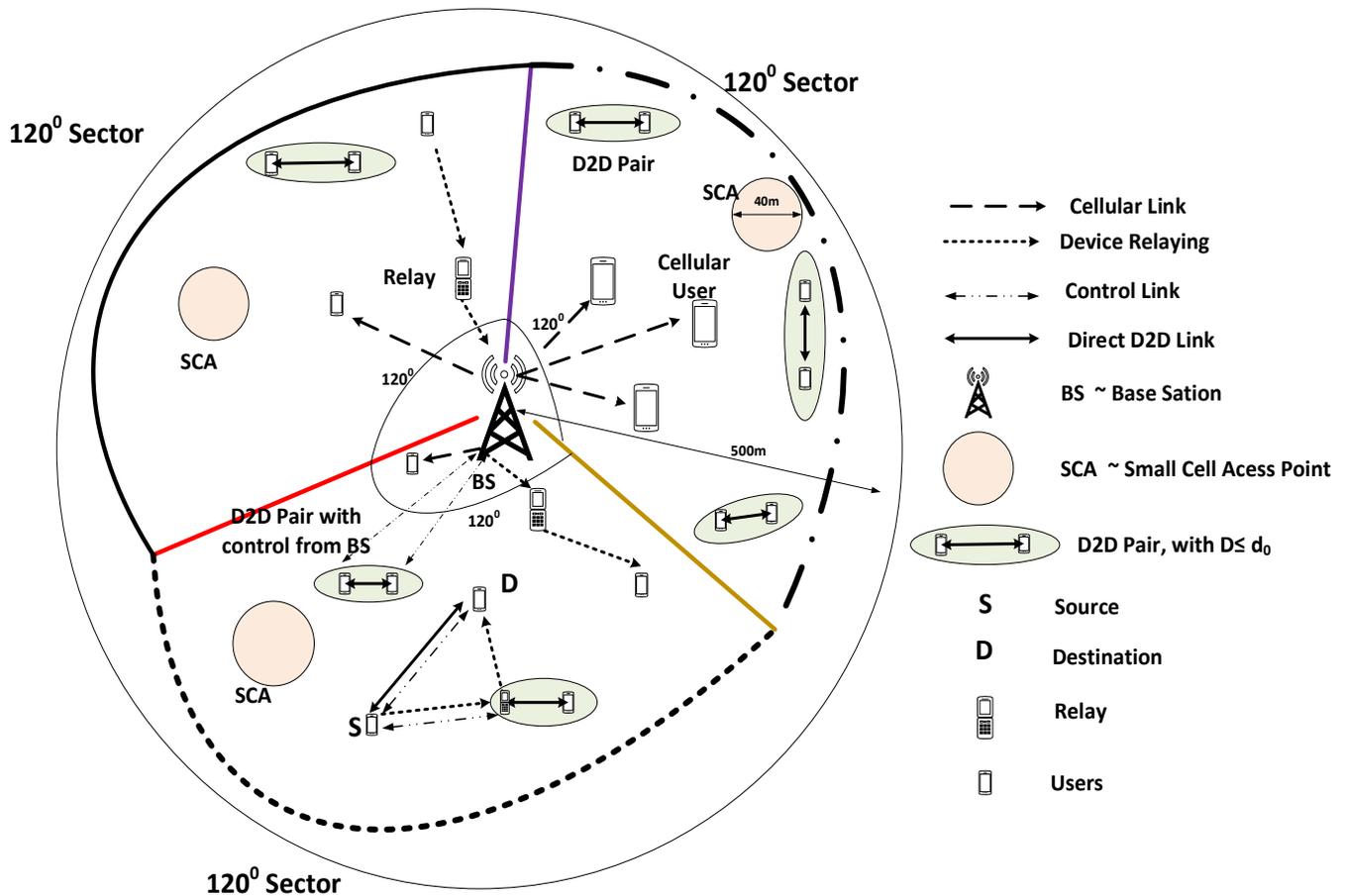

Fig.15. Proposed Architecture

With increasing number of subscribers and the rising demand for high data rates, the numbers of channels assigned to a cell become insufficient to support all the users. A need is felt to have a greater number of channels per unit coverage area of the cell. As a result, cell splitting is preferred in such scenarios. Splitting cells into sectors, with the use of directional antennas at the base station (BS) enhances their capability to handle more number of conversations at the same time. Cell sectoring is very useful for increasing the system subscriber handling capacity. Each sector then operates with its own set of frequency channels. With sectoring, the co-channel cell interference is greatly reduced, and considerable improvement in SINR is achieved. This is the reason for using sectored antennas at the base station, for the proposed model.

In the proposed model, a scenario with a single cell is considered. The cell is divided into three $120^0$ sectors, as is expected with the use of a sectored antenna. Each sector can have any number of users. The primary aim is to offload traffic from the base station and bring about an efficient device-to-device (D2D) communication with optimal resource allocation. The UEs that are close to the BS are served by the BS only, that is, those UEs will operate in the cellular mode. The UEs that are far from the BS, in congested areas or at the cell edge and are in proximity, communicate through D2D links. The formation of direct links between user equipments (UEs) is dependent on the distance between them, which must be less than or equal to the threshold distance, $d_0$. Thus, D2D link formation occurs when distance between any two UEs, $m$ and $n$ is such that $d(m, n) \leq d_0$. When once the distance constraint is met, the UE acts as a cellular user (CUE) or D2D user (DUE). The architecture aims at maximizing throughput, minimizing latency, enhancing system capacity, and efficiently utilizing the licensed spectrum through optimal resource allocation.

### C. POWER CONTROL

Setting the optimum transmission power for reusing the frequency is an area of interest for the researchers. It is particularly important in case of uplink transmissions because of the near-far effect and co-channel interference. Once a maximum power level is allocated to the D2D users, then the Quality of Service (QoS) of the cellular users is maintained in the network. Controlling power effectively mitigates interference in cellular networks. For D2D under laid cellular networks, there has been a considerable interest in power control methods. A limit is set upon the power level of the D2D transmitter and its reuse partner (the cellular user), in order to maximize the overall system throughput. This is expressed as

$$(P_i^{CU}, P_j^{DD}) = \arg max_{P_i^{CU}, P_j^{D2D}} \log_2(1 + \delta_{cu(i)})$$
$$+ \log_2(1 + \delta_{d(j)}) \quad (15)$$

subject to (12), (13), along with (16) and (17), given as

$$P_i^{CU} \leq P_{CU}^m \quad (16)$$
$$P_j^{D2D} \leq P_m^{DD} \quad (17)$$

where $P_{CU}^m$ and $P_m^{DD}$ set the maximum limits to the power of the cellular user and D2D transmitter, respectively.

To regulate the SINR degradation of cellular users, statistical power control schemes have been discussed in [72], [73], for different channel models. Some power control techniques are introduced in [74] and [75], through which improved performance can be achieved. A new distributed scheme for power control has been proposed in [76] in which a D2D underlay scenario is considered. The technique aims at minimization of the overall power consumption of the network, considering the optimal SINR target which is achieved with the use of Augmented Lagrangian Penalty Function (ALPF) method. Solving of equation (17) needs accessible full channel matrix. An algorithm with low complexity has been proposed in [77], based on game-theory, for selection of source and controlling power. It uses stackelberg game model to show the impact of improvement in D2D transmission quality.

[78] develop centralized and distributed algorithms for power control in a D2D network underlaying cellular network. A near optimal scheme for power control or rate control depending upon the condition of the channel is proposed in [79], thus reducing computational complexity. For the maximization of energy efficiency in the network, the authors of [80] propose an iterative joint resource allocation and power control technique. A penalty function approach adopted. In order to improve the quality of D2D communication underlaying cellular networks, an auction based power allocation approach is investigated in [81]. It is a low complexity algorithm, using a reverse iterative combinatorial auction and provides high system efficiency. Many other power control algorithms exist in literature. There is still ongoing research in this context as controlling power levels is essential for managing interference between the D2D users and cellular users.

### D. INTERFERENCE MANAGEMENT

Enabling D2D links within a cellular network pose a big threat of interference to the cellular links in the network. D2D links can cause interference between cellular users and D2D users, resulting in an increase in intra-cell interference. Inter-cell interference is also possible with D2D communication underlaying cellular communication. Interference can be mitigated through mode selection, optimum resource allocation, power control. Setting the maximum transmit power limits of the D2D transmitter is an effective technique of limiting the interference between DUEs and CUEs. A general scenario of interference in D2D under laid cellular networks is depicted in Fig.16.

A very critical term related to interference avoidance is mode selection. Generally, distance between the D2D users and cellular users is considered for mode selection (Overlay/underlay) [74]. Also, distance between cellular user and the BS is an important parameter for selection of the mode in the network, thus avoiding interference. In [82], MIMO transmission schemes are introduced for interference avoidance, resulting in a great enhancement of D2D SINR.

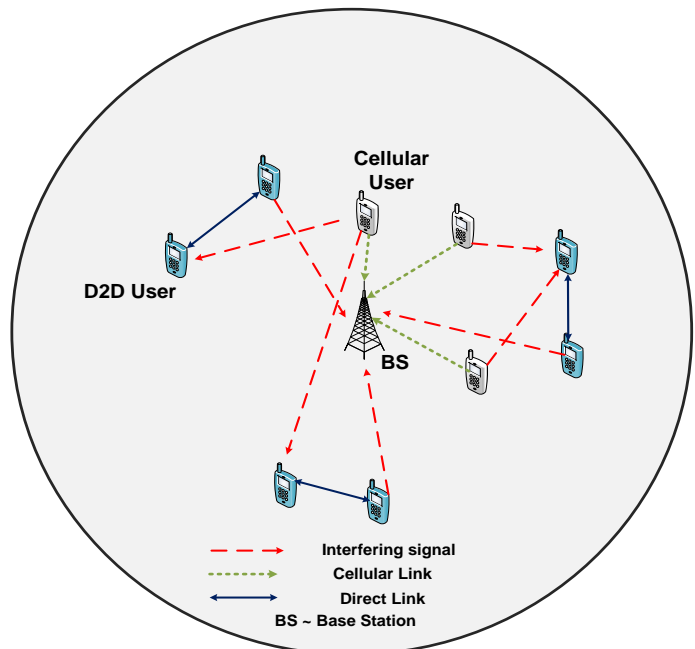

Fig.16.An interference scenario in D2D underlaid cellular network

Due to interfering signals, the received contain three components:

*Received Signal= Desired signal + Outside interference signal+D2D interference signal* (18)

Interference at the receiver must be minimized so that a higher value of SINR is achieved. This can be achieved by modulation and coding scheme (MCS), which supports error-free reception of information. The D2D interference signals can be reduced, but interference from outside sources is hard to avoid (equation (18)).

The authors of [83] take into consideration a D2D underlaying communication network for interference cancellation, along with the transmission powers for maximizing the utility of the network. Significant gains are enjoyed by the users in terms of spectral efficiency. In [84], authors propose a novel interference coordination scheme for improving system throughput and efficient resource utilization in a multicast D2D network. The authors of [85] concentrate on managing interference between D2D users and cellular users by discussing the range of an interference suppression area (ISA) which classifies the strength of the interference between the cellular and D2D users and influences the system performance. Adequate adjustment of the range of ISA can help achieve optimal system performance. Interference management using network coding is discussed in [86]. In a cellular system with users undergoing cellular communication, along with D2D multicast communication, both sharing the same spectrum, the interference scenarios are evaluated in [87]. Interference in such a scenario can be mitigated by power control, followed by optimal resource allocation. Thus, different approaches are adopted by different researchers for interference mitigation between D2D links and cellular links, and can be categorized as interference avoidance schemes; interference cancellation schemes, or interference coordination schemes, as shown in Fig.17. The authors in [88] provide a comprehensive survey on interference management in D2D communication.

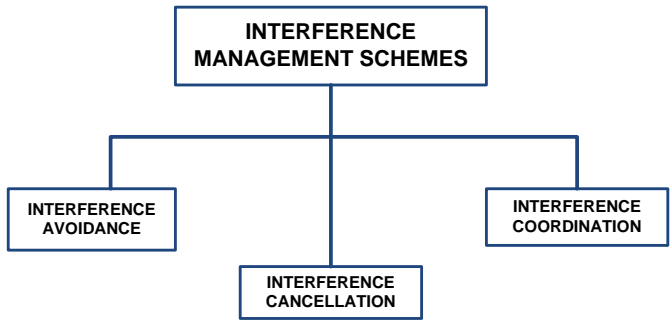

Fig. 17. Interference management schemes in D2D

*E. SECURITY*

Prior to the acceptance and implementation of the D2D technique in cellular network, security needs to be well addressed. The channels are vulnerable to a number of security attacks like eavesdropping, message modification, and node impersonation. To prevent these, cryptographic solutions can be used to encrypt the information before transmission. The security schemes provided by the cellular operators can be used by the D2D users if they are under their coverage. But, users outside the coverage of the operators can't be secured. In this case, security signals may be passed on through relays. Since relays are highly susceptible to malicious attacks, like eavesdropping attack, free riding attack, denial of service attack [89]. Thus, designing security schemes for D2D communication is an important challenge to be addressed.

To make D2D communication secure, physical layer security plays a key role [90]. Incorporating security features in D2D communication, at the physical layer is beneficial. For providing physical layer security, the received SINR at the eavesdropper need to be minimized. Beam forming techniques enhance security in cellular networks. Under a typical attack on the ongoing D2D communication, if Alice transmits some information '*x*' to the Bob, and it is captured by the intruder, then the signal received at the bob is

$$Y_{bob} = \sqrt{P_{alice}}\, H_{ab}\, x + \text{и}_b \quad (19)$$

and at the intruder is

$$Y_I = \sqrt{P_{alice}}\, H_{ae} x + \text{и}_i \quad (20)$$

where $Y_{bob}$, $Y_I$ correspond to the received signals at the bob and intruder, respectively; $P_{alice}$ is the transmit power of the Alice; $H_{ab}$, $H_{ae}$ are Gaussian random variables for modeling the scalar channels, and $\text{и}_b$, $\text{и}_i$ represent noise at the bob and intruder. These equations model attack on a single hop D2D communication in a cellular network. These are used for secrecy rate maximization in a D2D communication network. Presence of noise components in the received signal prevents desirable information from reaching the Bob. [16], [17], [18], [19], [20] discuss about the security concerns in D2D communication. A general architecture for securing D2D communication is discussed in [91]. Optimal power can be allocated to the Alice and Bob, to prevent eavesdropping. The communication overhead and the key generation time needs to be taken into account while designing the security algorithms.

To assure true benefits of D2D communication, the above listed issues need special attention of the research community.

## VI. APPLICATION AREAS OF D2D COMMUNICATION

In view of the current and future wireless traffic scenario, a number of use cases of device-to-device (D2D) communication have been proposed by the researchers. D2D communication can be carried out by direct link establishment between sender and receiver, or with D2D users acting as relays within the networks. The most important application of device-to-device (D2D) communication is cellular offloading [92], which results in an increased network capacity. Others applications include multicasting [63], video dissemination [93], and M2M communication [94]. M2M communications will be highly benefitted by D2D communication. It is technology-independent, unlike D2D communication, which is dependent upon the technology. In case of emergency communications (public safety communication), D2Dcommunication is expected to play an essential role. It has the ability to assure public protection and disaster relief (PPDR) and national security and public safety services (NSPS) [95]. For example, in case of a natural calamity, like an earthquake, conventional cellular networks can get damaged. In such a case, a wireless network can be setup between terminals, using D2D communication.

D2D communication, upon integration with Internet of Things (IoT) shall support important applications. This will result in a truly interconnected wireless network. A typical application for such a scenario is Vehicle-to-Vehicle (V2V) communication, in the Internet of Vehicles (IoV). This application is particularly important in case of collision avoidance systems, like in coordinating braking systems among the vehicles.

Other possible use cases of D2D communication are multiuser MIMO (MU-MIMO) enhancement, cooperative relaying, and virtual MIMO. With D2D communication, paired users can exchange the information about the channel status directly. In this way, channel status information can be fed by the terminals to base stations and improve the performance of MU-MIMO [96].

Various other use cases that can be supported by D2D communications, include location-aware services, social networking, smart grids [94], [97], e-health, smart city etc. Some use cases of D2D communication have also been addressed in [54]. Thus, a wide variety of applications are offered for the next generation networks by device-to-device (D2D) communication.

## VII. CONCLUSION

In this paper, an extensive survey on device-to-device (D2D) communication has been performed. This emerging technology is expected to solve the various tribulations of the mobile network operators (MNOs), efficiently satisfying all the demands of the subscribers. A complete overview about the different types of D2D communication and the supported architectures has been brought up. A number of features can be used in conjunction with D2D communication, to enhance the functionality of cellular networks. Some challenges related to the implementation of device-to-device (D2D) communication have been brought up in this survey, and various algorithms for dealing with them have been discussed. Architecture has been proposed in the survey, for optimal resource allocation to the D2D users underlaying cellular networks. This is important to ensure efficient communication in the existing cellular networks. Some use cases have been quoted, where D2D communication will play a crucial role. Thus, D2D communication is an integral technology of the future networks, motivating the researchers to overcome the associated challenges in order to completely take advantage of its utility.

APPENDIX

A. STANDARDIZATION ACTIVITIES FOR D2D

Device-to-device communication is widely being accepted by the mobile stakeholders and they believe it to be a big success in wireless technology. Qualcomm, LTE-A and IEEE 802.15.4g (SUN) are at present involved in the standardization activities of D2D communication over the licensed band.

IEEEE 802.15.4g was first released as an amendment to the low rate WPAN in April 2012. It supports three different modulation techniques, FSK, DSSS and OFDM. Maximum data rate supported is upto 200kbps with a maximum range of about 2-3 km. SUN is highly energy efficient, which is a very attracting feature. Other IEEE standards include IEEE 802.15.8 (for PHY/MAC specification of D2D) and IEEE 802.16n.

D2D communications with and without infrastructure are being studied by 3GPP. Proximity-based Services (ProSe) and Group Communicaton System Enablers for LTE (GCSE_LTE) are discussed in [4], [5] and [6]. D2D ProSe considers various aspects of D2D communication, including one-to-one, one-to-many and one-hop relay and also addresses switching of mode between D2D mode and cellular mode. ProSe includes working on identifying UEs in proximity (peer discovery, and establishing direct links between them, so as to enable communication between them either directly or through a locally routed path via the eNB.

Three different use cases are being studied by 3GPP that reflect the main market drivers for ProSe
  o  Local commercial advertisement: This sends advertisements to passing devices automatically.
  o  Network offloading: This helps in avoiding congestion in the network, by enabling traffic to pass through direct links.
  o  Public Safety Communication: In case of absence of network coverage, public safety communication is supported.

The study related to feasibility of ProSe started in 2011 by the 3GPP Technical Specification Group (TSG) Service and System Aspect (SA), and these also defined the technical requirements in 2012. In the documents of Release 12, TS22.278 and TS22.115, the technical specifications were written. In Release 13, Radio Access Network related activities of ProSe were expected to be included. Presently, the work on RAN1 (Radio Layer1) is in the middle of its feasibility stage along with compilation of proposals for solutions but also evaluation models (channel, traffic, mobility) in TR 36.843. The work in RAN2 (Radio Layer 2 and Layer 3-Radio Resource part) is in the first part of the feasibility stage. The work on CT1 (non-access stratum protocols) has not been started yet and will start on completion of the work in SA2.

TABLE VII
STANDARIZATION OF D2D COMMUNICATION

| ORGANIZATION | STANDARD |
| --- | --- |
| IEEE | IEEE 802.15.4 g (SUN) |
|  | IEEE 802.15.8 |
|  | IEEE 802.16n |
| QualComm | FlashLinQ |
| 3GPP | ProSe (Release 12) |

Technical specifications are provided in [98] in which GCSE_LTE refers to the 3GPP architecture based content distribution mechanism and expect the support of an efficient and fast communication. Intense research activities and meetings have been organized and still being organized by the 3GPP TSG RAN WG1. In spite of the ongoing debates, a lot of work on D2D communication has been done.

Initially, D2D communication was proposed in academia for enabling multihop relays in the cellular networks [99]. Later, it started gaining importance for various use cases and improving spectral efficiency. Qualcomm's FlashLinQ [3] was the first attempt towards D2D implementation. It is a PHY/MAC architecture which is needed for D2D underlaying networks. FlashLinQ is an efficient technique that provides peer discovery, synchronization of timing and management of link in D2D-enabled cellular networks.

Few researchers have proposed some protocols for device-to-device communicaton in cellular networks. Protocol stacks for inband and outband D2D communication have been introduced in [100] and [101], respectively. In [100], the modifications in terms of architecture and protocol have been given, that need to be made to the existing cellular networks. An important architectural modification involves addition of a D2D server in or out of the core network, along with suitable interfaces. In [101], the authors mainly aim at opportunistic relaying of packets. A performance evaluation of the D2D networks using existing simulators like OPNET [102], NS3 [103], Omnet++ [104] is expected to bring about fruitful results. Android softwares are also being developed and worked upon for D2D communication. A summary of the various activities is given in Table VII.

B. ONGOING PROJECTS ON D2D

Various ongoing D2D communication projects have been tabulated in Table VIII.

TABLE VIII
D2D RELATED ONGOING PROJECTS

| Research Project/Organsization | Objective |
| --- | --- |
| METIS D2D | Increasing coverage, offloading backhaul, improving spectrum usage, enabling new services |
| CODEC | Resource allocation in D2D communication |
| WiFiUS | D2D communication at millimeter frequencies |

## C. ABBREVIATIONS USED IN PAPER

Various abbreviations used in the paper have been listed in Table IX.

TABLE IX
LIST OF ABBREVIATIONS USED IN THE PAPER

| Abbreviation | Explanation |
|---|---|
| 3GPP | 3$^{rd}$ Generation Partnership Project |
| ACK/NACK | Acknowledgement/Negative Acknowledgement |
| ALPF | Augmented Lagrangian Penalty Function |
| ARQ | Automatic Repeat Request |
| BDMA | Beam Division Multiple Access |
| BS | Base Station |
| CDMA | Code Division Multiple Access |
| CRN | Cognitive Radio Network |
| CSI | Channel State Information |
| CUE | Cellular User Equipment |
| DSSS | Direct Sequence Spread Spectrum |
| DUE | D2D User Equipment |
| DVB | Digital Video Broadcasting |
| EDGE | Enhanced Data rate for GSM Evolution |
| eNB | Evolved Node B |
| EVDO | Evolution-Data Optimized |
| FBMC | Fiber Bank Mulitcarrier |
| GFSK | Gaussian Frequency Shift Keying |
| GPRS | General Packet Radio Service |
| GPS | Global Positioning System |
| HARQ | Hybrid Automatic Repeat Request |
| HOM | Handover Margin |
| HSUPA/HSDPA | High Speed Uplink/Downlink Packet Access |
| IoT | Internet of Things |
| IoV | Internet of Vehicles |
| IS-95 | Interim Standard-95 |
| ISA | Interference Suppression Area |
| ISM | Industrial, Scientific, Medical |
| ITU | International Telecommunication Union |
| LOS | Line Of Sight |
| LTE | Long Term Evolution |
| LTE-A | Long Term Evolution-Advanced |
| M2M | Machine-to-Machine |
| MANET | Mobile Ad-hoc Network |
| MIMO | Multiple Input Multiple Output |
| MMS | Multimedia Messaging Service |
| mmWave | Millimeter Wave |
| MNO | Mobile Network Operators |
| MU-MIMO | Multi-User MIMO |
| NGN | Next Generation Networks |
| NSPS | National Security and Public Safety Services |
| OFDMA | Orthogonal Frequency Division Multiple Access |
| PPDR | Public Protection and Disaster Relief |
| ProSe | Proximity Service |
| QoS | Quality of Service |
| RB | Resource Block |
| RSRP | Reference Signal Receiving Power |
| SC-FDMA | Single Carrier Frequency Division Multiple Access |
| SINR | Signal-to-Interference-plus-Noise-Ratio |
| TDS | Time Division Scheduling |
| TTT | Time to Trigger |
| UDN | Ultra Dense Network |
| UE | User Equipment |
| UMTS | Universal Mobile Telecommunications System |
| V2V | Vehicle-to-vehicle |
| VLC | Visible Light Communication |
| WCDMA | Wideband Code Division Multiple Access |
| WiMax | Wireless interoperability for microwave access |
| WLAN | Wireless Local Area Network |

REFERENCES

[1] D. Astely et al., "LTE Release 12 and Beyond." *IEEE Commun. Mag.*, vol.51, no. 7, 2013, pp. 154-60.

[2] Chai, Yingqi, Qinghe Du, and Pinyi Ren. "Partial time-frequency resource allocation for device-to-device communications underlaying cellular networks." *Communications (ICC), 2013 IEEE International Conference on*. IEEE, 2013.

[3] X. Wu, S. Tavildar, S. Shakkottai, T. Richardson, J. Li, R. Laroia, and A. Jovicic, "FlashLinQ: A synchronous distributed scheduler for peer-to-peer ad hoc networks", in IEEE Allerton Conference on Communication, Control and Computing, 2010, pp.514-521.

[4] 3GPP, "3rd generation partnership project; technical specification group SA; feasibility study for proximity services (ProSe) (Release 12)," TR 22.803 V12.2.0, June 2013.

[5] "3rd generation partnership project; technical specification group SA; study on architecture enhancements to support proximity services (ProSe) (Release 12)," TR 23.703 V2.0.0, March 2014.

[6] "3rd Generation Partnership Project; Technical Specification Group Services and System Aspects; Group Communication System Enablers for LTE (GCSE LTE) (Release 12)," TS 22.468 V12.0.0, June 2013.

[7] K. R. Santhi, V. K. Srivastava, G. SenthilKumaran, and A. Butare, ``Goals of true broad band's wireless next wave (4G_5G),'' in *Proc. IEEE 58$^{th}$ Veh. Technol. Conf.*, vol. 4. Oct. 2003, pp. 2317_2321.

[8] T. Halonen, J. Romero, and J. Melero, Eds., *GSM, GPRS and EDGE Performance: Evolution Towards 3G/UMTS*. NewYork,NY, USA:Wiley, 2003.

[9] Akhil Gupta, Rakesh Kumar Jha. "A survey of 5g network: Architecture and emerging technologies." *Access, IEEE* 3 (2015): 1206-1232.

[10] Cisco, V. N. I. "The Zettabyte Era: Trends and Analysis." *Updated (29/05/2013),http://www.cisco.com/c/en/us/solutions/collateral/servicep rovider/visual-networking-index-vni/VNI_Hyperconnectivity_WP. html* (2014).

[11] Bansal, Gaurav, et al. "Some research issues in cognitive radio networks."*AFRICON 2007*. IEEE, 2007.

[12] Cheng, Peng, et al. "Resource allocation for cognitive networks with D2D communication: An evolutionary approach." *Wireless Communications and Networking Conference (WCNC), 2012 IEEE*. IEEE, 2012.

[13] Tehrani, Mohsen Nader, Mustafa Uysal, and Halim Yanikomeroglu. "Device-to-device communication in 5G cellular networks: challenges, solutions, and future directions." *Communications Magazine, IEEE* 52.5 (2014): 86-92.

[14] Liu, Ziyang, et al. "Transmission capacity of D2D communication under heterogeneous networks with dual bands." *Cognitive Radio Oriented Wireless Networks and Communications (CROWNCOM), 2012 7th International ICST Conference on*. IEEE, 2012.

[15] Hakola, Sami, et al. "Device-to-device (D2D) communication in cellular network-performance analysis of optimum and practical communication mode selection." *Wireless Communications and Networking Conference (WCNC), 2010 IEEE*. IEEE, 2010.

[16] Cha, Inhyok, et al. "Trust in M2M communication." *Vehicular Technology Magazine, IEEE* 4.3 (2009): 69-75.

[17] Yue, Jianting, et al. "Secrecy-based access control for device-to-device communication underlaying cellular networks." *Communications Letters, IEEE* 17.11 (2013): 2068-2071.

[18] Perrig, Adrian, John Stankovic, and David Wagner. "Security in wireless sensor networks." *Communications of the ACM* 47.6 (2004): 53-57.

[19] Zhou, Yun, Yuguang Fang, and Yanchao Zhang. "Securing wireless sensor networks: a survey." *Communications Surveys & Tutorials, IEEE* 10.3 (2008): 6-28.

[20] Muraleedharan, Rajani, and Lisa A. Osadciw. "Jamming attack detection and countermeasures in wireless sensor network using ant system." *Defense and Security Symposium*. International Society for Optics and Photonics, 2006.

[21] Niu, Yong, et al. "Exploiting Device-to-Device Communications to Enhance Spatial Reuse for Popular Content Downloading in Directional mmWave Small Cells.", *IEEE Transaction on Vehicular Technology*, Vol PP, Issue:99, 2015

[22] Qiao, Jian, et al. "Enabling device-to-device communications in millimeter-wave 5G cellular networks." *Communications Magazine, IEEE* 53.1 (2015): 209-215.

[23] Singh, Sumit, et al. "Millimeter wave WPAN: Cross-layer modeling and multi-hop architecture." *INFOCOM 2007. 26th IEEE International Conference on Computer Communications. IEEE*. IEEE, 2007.

[24] Cao, Yang, Tao Jiang, and Chonggang Wang. "Cooperative device-to-device communications in cellular networks." *Wireless Communications, IEEE* 22.3 (2015): 124-129.

[25] Lefie Wang, Tao Peng, Yufeng Yang, and Wenbo Wang, "Interference constrained relay selection of d2d communication for relay purpose underlaying cellular networks," in *Wireless Communications, Networking and Mobile Computing (WiCOM), 8th international Conference on*, 2012, pp. 1-5

[26] Chen, Xu, et al. "Exploiting social ties for cooperative D2D communications: a mobile social networking case." *Networking, IEEE/ACM Transactions on* 23.5 (2015): 1471-1484.

[27] Del Carpio, Luis Felipe, et al. "Simple Clustering Methods for Multi-Hop Cooperative Device-to-Device Communication." *Vehicular Technology Conference (VTC Spring), 2015 IEEE 81st*. IEEE, 2015.

[28] Chen, Ho-Yuan, Mei-Ju Shih, and Hung-Yu Wei. "Handover Mechanism for Device-to-Device Communication.", *IEEE Conference on Standards for Communication and Networking,* 2015, 72-77

[29] Yilmaz, Osman NC, et al. "Smart mobility management for D2D communications in 5G networks." *Wireless Communications and Networking Conference Workshops (WCNCW), 2014 IEEE*. IEEE, 2014.

[30] Radwan, Ayman, and Jonathan Rodriguez, eds. *Energy Efficient Smart Phones for 5G Networks*. Springer, 2015.

[31] Mumtaz, Sami, et al. "Direct mobile-to-mobile communication: Paradigm for 5G." *Wireless Communications, IEEE* 21.5 (2014): 14-23.

[32] Du, Jinling, et al. "A compressed HARQ feedback for device-to-device multicast communications." *Vehicular Technology Conference (VTC Fall), 2012 IEEE*. IEEE, 2012.

[33] Miao, M. M., Jian Sun, and S. X. Shao. "Cross-layer optimization schemes based on HARQ for Device-to-Device communication." *Computing, Communications and IT Applications Conference (ComComAp), 2014 IEEE*. IEEE, 2014.

[34] 3GPP, "3GPP TSG Services and system aspects: Local IP access and selected IP traffic offload (LIPA-SIPTO)(Release 10)," 3GPP, Tech. Rep. TR 23.829, Aug, 2011.

[35] Yang, Mi Jeong, et al. "Solving the data overload: Device-to-device bearer control architecture for cellular data offloading." *Vehicular Technology Magazine, IEEE* 8.1 (2013): 31-39.

[36] Malandrino, Francesco, Claudio Casetti, and Carla-Fabiana Chiasserini. "Toward D2D-enhanced heterogeneous networks." *Communications Magazine, IEEE* 52.11 (2014): 94-100.

[37] Baldemair, Robert, et al. "Ultra-dense networks in millimeter-wave frequencies." *Communications Magazine, IEEE* 53.1 (2015): 202-208.

[38] Chen, Siyi, et al. "System-level performance evaluation of ultra-dense networks for 5G." *TENCON 2015-2015 IEEE Region 10 Conference*. IEEE, 2015.

[39] Ejaz, Waleed, Ghalib A. Shah, and Hyung Seok Kim. "Optimal Entropy-Based Cooperative Spectrum Sensing for Maritime Cognitive Radio Networks." *Entropy* 15.11 (2013): 4993-5011.

[40] Sakr, Ahmed Hamdi, et al. "Cognitive spectrum access in device-to-device-enabled cellular networks." *Communications Magazine, IEEE* 53.7 (2015): 126-133.

[41] Khoshkholgh, Mohammad G., et al. "Connectivity of Cognitive Device-to-Device Communications Underlying Cellular Networks." *Selected Areas in Communications, IEEE Journal on* 33.1 (2015): 81-99.

[42] Sakr, Ahmed Hamdi, and Ekram Hossain. "Cognitive and energy harvesting-based D2D communication in cellular networks: Stochastic geometry modeling and analysis." *Communications, IEEE Transactions on* 63.5 (2015): 1867-1880.

[43] Mumtaz, Shahid, et al. "Cognitive vehicular communication for 5G."*Communications Magazine, IEEE* 53.7 (2015): 109-117.

[44] Liu, Jiangchuan, et al. "Device-to-device communication in LTE-advanced networks: a survey.", IEEE Communications Surveys and Tutorials, Volumne 17, 2014.

[45] Wu, Yue, et al. "Network coding in device-to-device (D2D) communications underlaying cellular networks." *Communications (ICC), 2015 IEEE International Conference on*. IEEE, 2015.

[46] Pahlevani, Peyman, et al. "Novel concepts for device-to-device communication using network coding." *Communications Magazine, IEEE* 52.4 (2014): 32-39.

[47] Asadi, Arash, Qing Wang, and Vincenzo Mancuso. "A survey on device-to-device communication in cellular networks." *Communications Surveys & Tutorials, IEEE* 16.4 (2014): 1801-1819.

[48] W. ALLIANCE, "Wi-Fi Peer-to-Peer (P2P) Specification v1. 1, "WI-FI ALLIANCE SPECIFICATION, vol. 1, pp. 1-159, 2010.

[49] S. Bluetooth, "Bluetooth specification version 1.1," *Available HTTP: http://www.bluetooth.com, 2001.*

[50] Z. Alliance, "Zigbee specification," Document 053474r06, Version, vol. 1, 2006

[51] Lee, Woongsup, Juyeop Kim, and Sang-Won Choi. "New D2D Peer Discovery Scheme based on Spatial Correlation of Wireless Channel." (2016), IEEE Transaction on Vehicular Technology

[52] Feng, Daquan, et al. "Device-to-device communications in cellular networks." *Communications Magazine, IEEE* 52.4 (2014): 49-55.

[53] Fodor, Gábor, et al. "Design aspects of network assisted device-to-device communications." *Communications Magazine, IEEE* 50.3 (2012): 170-177.

[54] L. Lei *et al.*, "Operator Controlled Device-to-Device Communications in LTE-Advanced Networks," *IEEE Wireless Commun.*, vol. 19, no. 3, 2012, pp. 96–104

[55] Maciel, Tarcisio F., et al. "Network-assisted neighbor discovery based on power vectors for d2d communications." *Vehicular Technology Conference (VTC Spring), 2015 IEEE 81st*. IEEE, 2015.

[56] Zhang, Bentao, et al. "Social-aware peer discovery for D2D communications underlaying cellular networks." *Wireless Communications, IEEE Transactions on* 14.5 (2015): 2426-2439.

[57] Nguyen, Phong, et al. "Network-assisted device discovery for LTE-based D2D communication systems." *Communications (ICC), 2014 IEEE International Conference on*. IEEE, 2014.

[58] Tang, Huan, Zhi Ding, and Bernard C. Levy. "Enabling D2D Communications Through Neighbor Discovery in LTE Cellular Networks." *Signal Processing, IEEE Transactions on* 62.19 (2014): 5157-5170.

[59] Zou, Kingsley Jun, et al. "Proximity discovery for device-to-device communications over a cellular network." *Communications Magazine, IEEE* 52.6 (2014): 98-107.

[60] Phunchongharn, Phond, Ekram Hossain, and Dong In Kim. "Resource allocation for device-to-device communications underlaying LTE-advanced networks." *Wireless Communications, IEEE* 20.4 (2013): 91-100.

[61] M. R. Garey and D. S. Johnson, *Computers and* Intractability: A Guide to the Theory of NP-Completeness, W. H. Freeman and Co., 1979

[62] Yu, Chia-Hao, et al. "Resource sharing optimization for device-to-device communication underlaying cellular networks." *Wireless Communications, IEEE Transactions on* 10.8 (2011): 2752-2763.

[63] Zhou, Bin, et al. "Intracluster device-to-device relay algorithm with optimal resource utilization." *Vehicular Technology, IEEE Transactions on* 62.5 (2013): 2315-2326.

[64] Feng, Daquan, et al. "Device-to-device communications underlaying cellular networks." *Communications, IEEE Transactions on* 61.8 (2013): 3541-3551.

[65] Botsov, Mladen, et al. "Location dependent resource allocation for mobile device-to-device communications." *Wireless Communications and Networking Conference (WCNC), 2014 IEEE*. IEEE, 2014.

[66] Liu, Jia, et al. "A Resource Reuse Scheme of D2D Communication Underlaying LTE Network with Intercell Interference." *Communications and Network* 5.03 (2013): 187.

[67] A. Gjendemsjo, D. Gesbert, G. E. Oien, and S. G. Kiani, "Optimal power allocation and scheduling for two-cell capacity maximization," in *International Symposium on Modeling and Optimization in Mobile, Ad Hoc and Wireless Networks*, Apr. 2006.

[68] V. Chandrasekhar and Z. She, "Optimal uplink power control in two cell systems with rise-over-thermal constraints," *IEEE Commun. Let.*, vol. 12, no. 3, Mar. 2008.

[69] Wang, Bin, et al. "Resource allocation optimization for device-to-device communication underlaying cellular networks." *Vehicular Technology Conference (VTC Spring), 2011 IEEE 73rd*. IEEE, 2011.

[70] Chen, Biwei, Jun Zheng, and Yuan Zhang. "A time division scheduling resource allocation algorithm for D2D communication in cellular networks."*Communications (ICC), 2015 IEEE International Conference on*. IEEE, 2015.

[71] Zheng, Jun, Biwei Chen, and Yuan Zhang. "An Adaptive Time Division Scheduling Based Resource Allocation Algorithm for D2D Communication Underlaying Cellular Networks." *2015 IEEE Global Communications Conference (GLOBECOM)*. IEEE, 2015.

[72] Yu, Chia-Hao, et al. "On the performance of device-to-device underlay communication with simple power control." *Vehicular Technology Conference, 2009. VTC Spring 2009. IEEE 69th*. IEEE, 2009.

[73] Yu, Chia-Hao, et al. "Performance impact of fading interference to device-to-device communication underlaying cellular networks." *Personal, Indoor and Mobile Radio Communications, 2009 IEEE 20th International Symposium on*. IEEE, 2009.

[74] Si Wen, Xiaoyue Zhu, Zhesheng Lin, Xin Zhang, and Dacheng Yang,"Optimization of interference coordination schemes in device-to-device (d2d) communication," in Communications and Networking in China (CHINACOM), 2012 7th International ICST Conference on, 2012, pp. 542–547.

[75] Dongyu Wang and Xiaoxiang Wang, "An interference managment scheme for device-to-device multicast spectrum sharing hybrid network,"in Personal Indoor and Mobile Radio Communications (PIMRC), 2013 IEEE 24rd International Symposium on, 2013.

[76] Fodor, Gábor, and Norbert Reider. "A distributed power control scheme for cellular network assisted D2D communications." *Global Telecommunications Conference (GLOBECOM 2011), 2011 IEEE*. IEEE, 2011.

[77] Wang, Qijie, et al. "Quality-optimized joint source selection and power control for wireless multimedia D2D communication using stackelberg game." (2014).

[78] Lee, Namyoon, et al. "Power Control for D2D Underlaid Cellular Networks: Modeling, Algorithms, and Analysis." *Selected Areas in Communications, IEEE Journal on* 33.1 (2015): 1-13.

[79] Song, Hojin, et al. "Joint Power and Rate Control for Device-to-Device Communications in Cellular Systems." *Wireless Communications, IEEE Transactions on* 14.10 (2015): 5750-5762.

[80] Jiang, Yanxiang, et al. "Energy Efficient Joint Resource Allocation and Power Control for D2D Communications."*IEEE Transaction on Vehicular Technology* (2015).

[81] Xu, Chen, et al. "Subcarrier and power optimization for device-to-device underlay communication using auction games." *Communications (ICC), 2014 IEEE International Conference on*. IEEE, 2014.

[82] Jänis, Pekka, et al. "Interference-avoiding MIMO schemes for device-to-device radio underlaying cellular networks." *Personal, Indoor and Mobile Radio Communications, 2009 IEEE 20th International Symposium on*. IEEE, 2009.

[83] Zhou, Liang, Kalle Ruttik, and Olav Tirkkonen. "Interference Canceling Power Optimization for Device to Device Communication." *Vehicular Technology Conference (VTC Spring), 2015 IEEE 81st*. IEEE, 2015.

[84] Wang, Dongyu, Xiaoxiang Wang, and Yuan Zhao. "An interference coordination scheme for device-to-device multicast in cellular networks."*Vehicular Technology Conference (VTC Fall), 2012 IEEE*. IEEE, 2012.

[85] Bin Guo, Shaohui Sun, Qiubin Gao, " Interference Management for D2D Communications Underlaying Cellular Networks at Cell Edge ", ICWMC, 2015

[86] Wang, Shuang, et al. "A Novel Interference Management Scheme in Underlay D2D Communication."*Vehicular Technology Conference (VTC Fall), 2015 IEEE 82nd*. IEEE, 2015.

[87] Wang, Dongyu, Xiaoxiang Wang, and Yuan Zhao. "An interference coordination scheme for device-to-device multicast in cellular networks."*Vehicular Technology Conference (VTC Fall), 2012 IEEE*. IEEE, 2012.

[88] Noura, Mahda, and Rosdiadee Nordin. "A Survey on Interference Management for Device-to-Device (D2D) Communication and its Challenges in 5G Networks." *Journal of Network and Computer Applications* (2016).

[89] Osanaiye, Opeyemi, Kim-Kwang Raymond Choo, and Mqhele Dlodlo. "Distributed Denial of Service (DDoS) Resilience in Cloud: Review and Conceptual Cloud DDoS Mitigation Framework." *Journal of Network and Computer Applications* (2016).

[90] Zhu, Daohua, et al. "Device-to-device communications: The physical layer security advantage." *Acoustics, Speech and Signal Processing (ICASSP), 2014 IEEE International Conference on*. IEEE, 2014.

[91] Wang, Mingjun, and Zheng Yan. "Security in D2D Communications: A Review." *Trustcom/BigDataSE/ISPA, 2015 IEEE*. Vol. 1. IEEE, 2015.

[92] X. Bao, U. Lee, I. Rimac, and R. R. Choudhury, "DataSpotting: offloading cellular traffic via managed device-to-device data transfer at data spots," *ACM SIGMOBILE Mobile Computing and CommunicationsReview*, vol. 14, no. 3, pp. 37–39, 2010.

[93] N. Golrezaei, A. F. Molisch, and A. G. Dimakis, "Base-station assisted device-to-device communications for high-throughput wireless video networks," in *Proceedings of IEEE ICC*, 2012, pp. 7077–7081

[94] N. K. Pratas and P. Popovski, "Low-rate machine-type communication via wireless device-to-device (D2D) links," *arXiv preprint arXiv:1305.6783*, 2013.

[95] Fodor, Gábor, et al. "Device-to-device communications for national security and public safety." *Access, IEEE* 2 (2014): 1510-1520.

[96] J. C. Li, M. Lei, and F. Gao, "Device-to-device (D2D) communication in MU-MIMO cellular networks," in *Proceedings of IEEE GLOBECOM*, 2012, pp. 3583–3587.

[97] Fey, Simon, et al. "Device-to-device communication for Smart Grid devices."Innovative Smart Grid Technologies (ISGT Europe), 2012 3rd IEEE PES International Conference and Exhibition on. IEEE, 2012.

[98] "3rd Generation Partnership Project; Technical Specification Group Services and System Aspects; Group Communication System Enablers for LTE (GCSE LTE) (Release 12)," TS 22.468 V12.0.0, June 2013.

[99] Lin, Ying-Dar, and Yu-Ching Hsu. "Multihop cellular: A new architecture for wireless communications." *INFOCOM 2000. Nineteenth Annual Joint Conference of the IEEE Computer and Communications Societies. Proceedings. IEEE*. Vol. 3. IEEE, 2000.

[100] B. Raghothaman, E. Deng, R. Pragada, G. Sternberg, T. Deng, and K. Vangannuru, "Architecture and protocols for LTE-based device-to-device communication" , in International Conference on Computing, Networking and Communication (ICNC), IEEE, 2013. Pp. 895-899

[101] "WiFi Direct and LTE D2D in action," Accepted for publication in IEEE Wireless Days, 2013

[102] OPNET. [Online]. Available: http://www.opnet.com/

[103] OPNET. [Online]. Available: http://www.opnet.com/

[104] A. Varga, "Omnet++ simulator," *Omnet++ simulator available at http://www. omnetpp. org*, 2007

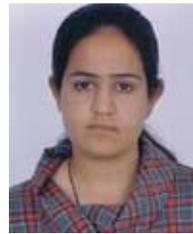

**Miss. Pimmy Gandotra (S'16)** received the B.E. degree in Electronics and Communication Engineering from Jammu University, Jammu and Kashmir, India, in 2013. She is currently pursuing the M.Tech degree in Electronics and Communication Engineering at Shri Mata Vaishno Devi University, Katra, Jammu and Kashmir, India

Her research interest includes the emerging technologies of 5G wireless communication network. Currently she is doing her research work on Resource Allocation in Device to Device communication. She is working on Qualnet simulation and Matlab tools for Wireless Communication.

She is receiving the teaching assistantship from MHRD. She is a student member of Institute of Electrical and Electronics Engineers (IEEE).

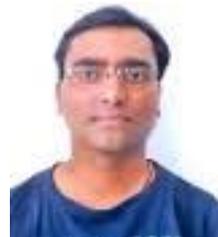

**Dr. Rakesh K Jha (S'10, M'13)** is currently an Assistant Professor in School of Electronics and Communication Engineering, Shri Mata Vaishno Devi University, katra, Jammu and Kashmir, India. He is carrying out his research in wireless communication, power optimizations, wireless security issues and optical communications. He has done B.Tech in Electronics and Communication Engineering from Bhopal, India and M.Tech from NIT Jalandhar, India. Received his PhD degree from NIT Surat, India in 2013.

He has published more than 30 International Journal papers and more than 20 International Conference papers. His area of interest is Wireless communication, Optical Fiber Communication, Computer networks, and Security issues.

Dr. Jha's one concept related to router of Wireless Communication has been accepted by ITU (International Telecommunication Union) in 2010. He has received young scientist author award by ITU in Dec 2010. He has received APAN fellowship in 2011 and 2012, and student travel grant from COMSNET 2012. He is a senior member of IEEE, GISFI and SIAM, International Association of Engineers (IAENG), ACCS (Advance Computing and Communication Society) and Association for Computing Machinery (ACM).